\documentclass[twocolumn,aps,prd,floatfix,preprintnumbers,a4paper,nofootinbib,superscriptaddress]{revtex4-2}
 
\usepackage{graphicx}				
\usepackage{placeins}
\usepackage{amssymb}
\usepackage{amsmath}
\usepackage[dvipsnames]{xcolor}
\usepackage{verbatim}
\usepackage{rotating}
\usepackage{comment}
\usepackage[colorlinks]{hyperref}

\usepackage{graphicx}
\usepackage{amsfonts}
\usepackage{dcolumn}
\usepackage{longtable}
\usepackage[caption=false]{subfig}
\usepackage{soul}
\usepackage{mathrsfs}
\usepackage{listings}
\usepackage{multirow}

\definecolor{LinkColor}{rgb}{0, 0, 0.75}
\definecolor{CiteColor}{rgb}{0.75, 0, 0}
\definecolor{UrlColor}{rgb}{0, 0, 0.75}
\hypersetup{linkcolor=LinkColor}
\hypersetup{citecolor=CiteColor}
\hypersetup{urlcolor=UrlColor}

\begin{document}

% Input macros (commands, shorthand and functions)
% definitions

\newenvironment{aside}
    {\begin{addmargin}[1em]{2em}% 2em left, 1em right
\begin{center} \noindent\rule{0.65\paperwidth}{0.4pt} \end{center}
    }
    {
    \begin{center} \noindent\rule{0.65\paperwidth}{0.4pt} \end{center}
\end{addmargin}
    }

% CUSTOM COMMANDS
\newcommand\xquote[1]{``#1"}
\newcommand\psinr[2]{$\psi_{{#1}{#2}}^{NR}$}
\newcommand\red[1]{{\color[rgb]{0.75,0.0,0.0} #1}}
\newcommand\redst[1]{\red{\st{#1}}}
\newcommand\bred[1]{{\color[rgb]{0.75,0.0,0.0} \textbf{#1}}}
\newcommand\green[1]{{\color[rgb]{0.0,0.60,0.08} #1}}
\newcommand\blue[1]{{\color[rgb]{0,0.20,0.65} #1}}
\newcommand\cyan[1]{{\color[HTML]{00c3ff} #1}}
\newcommand\bluey[1]{{\color[rgb]{0.11,0.20,0.4} #1}}
\newcommand\gray[1]{{\color[rgb]{0.7,0.70,0.7} #1}}
\newcommand\grey[1]{{\color[rgb]{0.7,0.70,0.7} #1}}
\newcommand\white[1]{{\color[rgb]{1,1,1} #1}}
\newcommand\darkgray[1]{{\color[rgb]{0.3,0.30,0.3} #1}}
\newcommand\orange[1]{{\color[rgb]{.86,0.24,0.08} #1}}
\newcommand\purple[1]{{\color[rgb]{0.45,0.10,0.45} #1}}
\newcommand\note[1]{\colorbox[rgb]{0.85,0.94,1}{\textcolor{black}{\textsc{\textsf{#1}}}}}

\newcommand*{\figfactor}{0.495}

% Math symbols
% ###################################### %
%\def\mh{{\hat{h}}} % model  for strain
\def\m#1{{\hat{#1}}} % model  for strain
\def\ma#1{{\hat{#1}_{\bm{a}}}} % model  for strain
\newcommand{\var}[1]{\mathcal{{#1}}}
\newcommand{\mlam}{{\bm{\lambda}}}
\newcommand{\lam}{{\lambda}}
\newcommand{\mLam}{{\bm{\Lambda}}}
\newcommand{\bigo}[1]{{\cal O}({#1})}
\newcommand{\braket}[2]{ {\langle {#1} \, | \, {#2} \rangle} }
\newcommand{\bra}[1]{ \langle {#1} |  }
\newcommand{\ket}[1]{ | {#1} \rangle }
\newcommand{\ketbra}[2]{ \ket{#1}\bra{#2} }
\newcommand{\tx}[1]{\text{#1}}
\def\T{\dagger}
% ###################################### %

% ###################################### %
\definecolor{lightblue}{rgb}{.82,.88,0.95}
\definecolor{lightred}{rgb}{0.95,.86,0.86}
\definecolor{yellow}{rgb}{0.95,0.95,0.86}
\definecolor{green}{rgb}{.90,1,0.95}
\definecolor{lightpurple}{rgb}{.95,0.85,0.95}
% ###################################### %

% CUSTOM DEFINITIONS
% ***************************************** %
\def\prd{Phys.Rev.D}
% ***************************************** %
\def\gt{Georgia Tech}
% ***************************************** %
\def\tk{{Teukolsky}}
% ***************************************** %
\def\mpl{multipole}
% ***************************************** %
\def\ee{Einstein's equations}
% ***************************************** %
\def\toolkit#1{NRDA--Toolkit{#1}}
% \def\toolkit#1{Data Analysis Toolkit{#1}}
% ***************************************** %
\def\wf#1{waveform#1}
% ***************************************** %
% \def\gr#1{General Relativity#1}
\def\gr#1{General Relativity#1
  (GR#1)\gdef\gr{GR}}
% ***************************************** %
\def\gwa#1{Gravitational Wave Astrophysics#1}
% ***************************************** %
\def\gwf#1{gravitational waveform#1}
% ***************************************** %
\def\gwa#1{\gw{} astronomy#1}
% ***************************************** %
\def\grad#1{gravitational radiation#1}
% ***************************************** %
\def\nht#1{No-Hair Theorem#1}
% ***************************************** %
\def\rm#1{\mathrm{#1}}
% ***************************************** %

% Referencing
% ***************************************** %
\def\prt#1{Part~(\ref{#1})}
% ***************************************** %
\def\Apx#1{Appendix~(\ref{#1})}
% ***************************************** %
\def\apx#1{Appx.~(\ref{#1})}
% ***************************************** %
\def\capx#1{Appx.~\ref{#1}}
% \def\ap#1{Appendix~(\ref{#1})}
% ***************************************** %
\def\ch#1{Chapter~(\ref{#1})}
% ***************************************** %
\def\cch#1{Chapter~\ref{#1}}
% ***************************************** %
\newcommand{\chs}[2]{Chapters~(\ref{#1}-\ref{#2})}
% ***************************************** %
\newcommand{\cchs}[2]{Chapters~\ref{#1}-\ref{#2}}
% ***************************************** %
\def\Sec#1{Section~\ref{#1}}
% ***************************************** %
\def\sec#1{Sec.~\ref{#1}}
% \def\sec#1{Sec.~(\ref{#1})}
% ***************************************** %
\def\csec#1{Sec.~\ref{#1}}
% ***************************************** %
\newcommand{\csecs}[2]{Secs.~\ref{#1}-\ref{#2}}
% ***************************************** %
\def\tk#1{Teukolsky#1}
% ***************************************** %
\newcommand{\secs}[2]{Secs.~\ref{#1}-\ref{#2}}
% ***************************************** %
\newcommand{\secsa}[2]{Sec.~\ref{#1} and Sec.~\ref{#2}}
% ***************************************** %
\def\Tbl#1{Table~(\ref{#1})}
% ***************************************** %
\def\tbl#1{Table~(\ref{#1})}
% ***************************************** %
\def\ctbl#1{Table~\ref{#1}}
% ***************************************** %
\def\Fig#1{Figure~\ref{#1}}
% ***************************************** %
\def\fig#1{Fig.~\ref{#1}}
% \def\fig#1{Fig.~(\ref{#1})}
% ***************************************** %
\def\cfig#1{Fig.~\ref{#1}}
% ***************************************** %
\newcommand{\figs}[2]{Figures~(\ref{#1}-\ref{#2})}
\newcommand{\Figs}[2]{Figures~(\ref{#1}-\ref{#2})}
\newcommand{\Figsa}[2]{Figures~(\ref{#1}) and (\ref{#2})}
% ***************************************** %
\def\Eqn#1{Equation~(\ref{#1})}
% ***************************************** %
\def\eqn#1{Eq.~(\ref{#1})}
% \def\eqn#1{\hyperref[#1]{Equation~\ref{#1}}} %
% ***************************************** %
\def\ceqn#1{Eq.~\ref{#1}}
% ***************************************** %
\newcommand{\Eqns}[2]{Equations~(\ref{#1}-\ref{#2})}
\newcommand{\Eqnsa}[2]{Equations~(\ref{#1}) and (\ref{#2})}
% ***************************************** %
\newcommand{\eqns}[2]{Eqs.~(\ref{#1}-\ref{#2})}
\newcommand{\eqnsa}[2]{Eqs.~(\ref{#1}) and (\ref{#2})}
% ***************************************** %
\newcommand{\ceqns}[2]{Eqs.~\ref{#1}-\ref{#2}}
\newcommand{\ceqnsa}[2]{Eqs.~\ref{#1} and \ref{#2}}
% ***************************************** %

% ***************************************** %
\def\tridiag#1{tridiagonal#1}
% ***************************************** %
\def\ply#1{{polynomial#1}}
% ***************************************** %
\def\plys{{polynomials}}
% ***************************************** %
\def\orthog{orthogonality}
% ***************************************** %
\def\da#1{data analysis#1}
% ***************************************** %
\def\sw{spin weighted}
% ***************************************** %
\def\swsh{spin weighted spherical harmonic}
% ***************************************** %
\def\lal#1{LIGO Analysis Library#1
  (LAL#1)\gdef\lal{LAL}}
% ***************************************** %
\def\nrda#1{\nr{} Data Analysis#1
  (NRDA#1)\gdef\nrda{NRDA}}
% ***************************************** %
\def\tt#1{\textit{transverse--traceless}#1
  (TT#1)\gdef\tt{TT}}
% ***************************************** %
\def\et#1{Einstein Telescope#1
  (ET#1)\gdef\et{ET}}
% ***************************************** %
\def\ce#1{Cosmic Explorer#1
  (CE#1)\gdef\ce{CE}}
% ***************************************** %
\def\ego#1{European Gravitational Observatory#1
  (EGO#1)\gdef\ego{EGO}}
% ***************************************** %
\def\lisa#1{Laser Interferometer Space Antenna#1
  (LISA#1)\gdef\lisa{LISA}}
% ***************************************** %
\def\ligo#1{Laser Interferometer Gravitational Wave Observatory#1
  (LIGO#1)\gdef\ligo{LIGO}}
% \def\ligo#1{LIGO#1}
% ***************************************** %
\def\igwn#1{International Gravitational Wave Network#1
  (IGWN)\gdef\igwn{IGWN}}
% ***************************************** %
\def\lvk#1{LIGO-Virgo-Kagra collaboration#1
  (LVK)\gdef\lvk{LVK}}
% ***************************************** %
\def\lv#1{#1
(LIGO-Virgo#1)\gdef\lv{LV}}
% ***************************************** %
\def\virgo#1{Virgo#1}
% ***************************************** %
\def\aligo#1{Advanced LIGO#1
  (Adv. LIGO#1)\gdef\aligo{Adv. LIGO}}
% ***************************************** %
% \def\snr#1{signal to noise ratio#1}
\def\snr#1{signal to noise ratio#1
  (SNR#1)\gdef\snr{SNR}}
% ***************************************** %
\def\psd#1{power spectral density#1
  (PSD#1)\gdef\psd{PSD}}
% ***************************************** %
\def\rom#1{reduced order model#1
  (ROM#1)\gdef\rom{ROM}}
% ***************************************** %
\def\gatech#1{Georgia Institute of Technology#1
  (GaTech#1)\gdef\gatech{GaTech}}
% ***************************************** %
\def\ffi#1{Fixed-Frequency Integration#1
  (FFI#1)\gdef\ffi{FFI}}
% ***************************************** %
\def\sxs#1{Simulating Extreme Spacetimes#1
  (SXS#1)\gdef\sxs{SXS}}
% ***************************************** %
\def\bam#1{Bifunctional Adaptive Mesh#1
  (BAM#1)\gdef\bam{BAM}}
% ***************************************** %
\def\adm#1{Arnowitt-Deser-Misner
	(ADM#1)\gdef\adm{ADM}}
\def\frmse#1{Fractional Root-Mean Square Error
	(FRMSE)\gdef\frmse{FRMSE}}
% ***************************************** %
\def\natpolys#1{\bh{} natural polynomials#1}
\def\Bhp#1{Black hole polynomials#1}
\def\bhp#1{black hole polynomials#1
	(BH polynomial#1)\gdef\bhp{BH polynomial#1}}
% ***************************************** %
\def\pj#1{Pollaczek–Jacobi#1}
\def\pjp#1{\pj{} polynomial#1}
\def\bio#1{biorthogonal#1}
% ***************************************** %
\def\bh#1{black hole#1
 (BH#1)\gdef\bh{BH}}
% \def\bh#1{black hole#1}
% ***************************************** %
\def\bbh#1{binary black hole#1
 (BBH#1)\gdef\bbh{BBH}}
% \def\bbh#1{binary \bh{}#1}
% ***************************************** %
\def\bhb#1{\bh{} binary#1}
% ***************************************** %
\def\slt#1{Sturm-Liouville theory#1}
% ***************************************** %

% ***************************************** %
\def\qnm#1{Quasi-Normal Mode#1
(QNM#1)\gdef\qnm{QNM}}
\def\Qnm#1{Quasi-Normal Mode#1}
\def\Qnms{Quasi-Normal Modes}
% ***************************************** %
\def\eob#1{Effective One Body#1
  (EOB#1)\gdef\eob{EOB}}
% ***************************************** %
\def\sws#1{spheroidal harmonics of spin weight -2#1}
\def\swy#1{spherical harmonics of spin weight -2#1}
% ***************************************** %
\def\gw#1{gravitational wave#1}
% \def\gw#1{gravitational wave#1
%  (GW#1)\gdef\gw{GW}}
% ***************************************** %
\def\gwa#1{gravitational wave astronomy#1}
% ***************************************** %
% \def\pn#1{Post-Newtonian#1}
\def\pn#1{Post-Newtonian#1
 (PN#1)\gdef\pn{PN}}
\def\pnl#1{post-Newtonian-like#1
  (PN-like#1)\gdef\pnl{PN-like}}
% ***************************************** %
% \def\nr{Numerical Relativity}
\def\NR{{\text{NR}}}
\def\nr{Numerical Relativity
 (NR)\gdef\nr{NR}}
% ***************************************** %
\def\pt{\bh{} perturbation theory}
% ***************************************** %
\def\GOLS#1{\textit{greedy ordinary least-squares}#1}
% ***************************************** %
\def\rd{ringdown}
% ***************************************** %
\def\imr{inspiral-merger-ringdown}
% \def\imr#1{inspiral-merger-ringdown#1
%   (IMR#1)\gdef\imr{IMR}}
% ***************************************** %
\def\cbc#1{compact object coalescence#1}
% ***************************************** %
\def\bbc#1{\bbh{} coalescence#1}
%\def\bbc#1{binary black hole coalescence#1
%  (BBC#1)\gdef\bbc{BBC}}
% ***************************************** %
\def\pc#1{principle component#1}
% ***************************************** %
\def\pca#1{principle component analysis#1
  (PCA#1)\gdef\pca{PCA}}
% ***************************************** %
\def\svd#1{Singular Value Decomposition#1
  (SVD#1)\gdef\svd{SVD}}
% ***************************************** %
\def\gs#1{Gram-Schmidt#1}
% ***************************************** %
\def\bhp#1{black hole polynomial#1}
\newcommand{\y}[2]{ {_{#1}}Y_{#2} }
\def\sylm{ \y{s}{\ell m } }
% ***************************************** %
\def\mo{\mathcal{D}}
\def\adjmo{{\mathcal{D}}^{\dagger}}
% ***************************************** %
\def\yo{\mathcal{D}_Y}
\def\adjyo{{\mathcal{D}_Y}^{\hspace{-2pt}\dagger}}
% ***************************************** %
\def\ro{\mathcal{D}_R}
\def\adjro{{\mathcal{D}_R}^{\hspace{-2pt}\dagger}}
% ***************************************** %
\def\bo{\mathcal{D}_B}
\def\adjbo{{\mathcal{D}_B}^{\hspace{-2pt}\dagger}}
% ***************************************** %
\def\so{\mathcal{D}_S}
\def\adjso{{\mathcal{D}_S}^{\hspace{-2pt}\dagger}}
% ***************************************** %
\def\to{\mathcal{D}_T}
\def\adjto{{\mathcal{D}_T}^{\hspace{-2pt}\dagger}}
% ***************************************** %
\def\adj#1{{#1}^{\dagger}}
\def\dadj#1{{#1}^{\ddag}}
% ***************************************** %
\def\ethp{\eth}
\def\ethm{{\eth'}}
\def\A{{\L m}}
\def\a{{\alpha}}
\def\mcl{\mathcal{L}}
\def\mct{{\mathcal{T}}}
\def\mcv{{\mathcal{V}}}
\def\cmcv{{\mcv^*}}
\def\cmct{{\mct^*}}
\def\tmct{{\tilde{\mct}}}
\def\tmcv{{\tilde{\mcv}}}
\def\mclo{{\mathcal{L}_o}}
\def\mcto{{\mathcal{T}_o}}
\def\mcvo{{\mathcal{V}_o}}
\def\cmcvo{{{\mcv_o^*}}}
\def\cmcto{{{\mct_o^*}}}
\def\tmcto{{\tilde{\mct}_o}}
\def\tmcvo{{\tilde{\mcv}_o}}
\def\tmcl{\tilde{\mcl}}
\def\mcp{\mathcal{P}}
\def\mcq{\mathcal{Q}}
\def\amcl{\adj{\mcl}}
\def\cmcl{{\mcl^*}}
\def\sjk{{\sigma_{\lp\ell}}}
\def\mcD{\mathcal{D}}
\def\mcL{\mathcal{L}}
\def\mcT{\mathcal{T}}
\def\mcV{\mathcal{V}}
\def\mcP{\mathcal{P}}
\def\mcQ{\mathcal{Q}}
\def\Lo{ {\mathcal{K}} }
\def\I{{\mathbb{I}}}
\def\max{\mathrm{max}}

% ***************************************** %
\def\gs{Gram-Schmidt}
\def\polys{polynomials}
% ***************************************** %

% ----------------------------------------- %
\def\lMn{{{\ell \M n}}}
\def\lmn{{{\ell m n}}}
\def\lmin{{{\ell_\mathrm{min}}}}
\def\lpmn{{{\ell' m n}}}
\def\lpmnp{{{\ell' m n'}}}
\def\lm{{{\ell m}}}
\def\lpm{{{\ell' m}}}
\def\l{{{\ell}}}
\def\n{{\bar{n}}}
\def\lmbn{{{\ell m \n}}}
\def\lpmbn{{{\ell' m \n}}}
\def\lp{{{\ell'}}}
\def\pp{p}

\mathchardef\minus = "002D
\newcommand{\swY}[4][]{{}_{{}_{#2}}\!Y^{#1}_{#3}(#4)}
\newcommand{\swSH}[5][]{{}_{{}_{#2}}S^{#1}_{#3}(#4;#5)}
\newcommand{\swS}[5][]{{}_{{}_{#2}}S^{#1}_{#3}(#4;#5)}
\newcommand{\scA}[4][]{{}_{{}_{#2}}A^{#1}_{#3}(#4)}
\newcommand{\YSH}[3][]{\mathcal{A}^{#1}_{#2}(#3)}

\def\LMaster{ \mathcal{L}_{t r \theta \phi} }
\def\LMasterB{ \mathcal{L}_{t r u \phi} }
\def\rp{ r_{+} }
\def\rm{ r_{-} }

\def\rhs{right hand side}
\def\lhs{left hand side}
\def\te{\tk{'s} equation}

\def\L{\bar{\ell}}
\def\M{\bar{m}}
\def\LM{{\L\M}}
\def\Lop{\mathcal{L}_{\k}}

% ***************************************** %
\newcommand{\brak}[2]{ \braket{#1}{#2} }
% ***************************************** %
%
% ***************************************** %
\newcommand*{\factor}{0.95} % for figure scale
\newcommand*{\rscale}{1.3}
% ***************************************** %

\newcommand{\hlgreen}[1]{\sethlcolor{green}\hl{#1}{\sethlcolor{yellow}}}
\newcommand{\hlyellow}[1]{\sethlcolor{yellow}\hl{#1}{\sethlcolor{yellow}}}
\newcommand{\hlblue}[1]{\sethlcolor{lightblue}\hl{#1}{\sethlcolor{yellow}}}
\newcommand{\hlred}[1]{\sethlcolor{lightred}\hl{#1}{\sethlcolor{yellow}}}
\newcommand{\hlpurple}[1]{\sethlcolor{lightpurple}\hl{#1}{\sethlcolor{yellow}}}

\newcommand{\qnms}{\qnm{s}}

\def\check#1{\red{#1}}
\def\new#1{\blue{#1}}
\def\remove#1{\hlred{#1}}
\newcommand{\cw}{\tilde{\omega}}
\newcommand{\CW}{\tilde{\Omega}}
\newcommand{\CWr}{{\Omega}^{\mathrm{r}}}
\newcommand{\CWc}{{\Omega}^{\mathrm{c}}}
\newcommand{\SC}{\mathcal{K}}
\newcommand{\CC}{\mathcal{C}}
\newcommand{\SCr}{\mathcal{K}^{\mathrm{r}}}
\newcommand{\SCc}{\mathcal{K}^{\mathrm{c}}}
\newcommand{\lalapprox}{\texttt{MMRDNS}}
\def\jf{j_f}
\def\mf{M_f}
\newcommand{\LL}{\bar{l}}
\newcommand{\MM}{\bar{m}}
\def\gmvp#1{greedy-multivariate-polynomial#1
  (\texttt{GMVP}#1)\gdef\gmvp{\texttt{GMVP}}}
\def\gmvr#1{greedy-multivariate-rational#1
  (\texttt{GMVR}#1)\gdef\gmvr{\texttt{GMVR}}}

% Paper ref macros
\def\PaperOne{\hyperlink{cite.London:2023aeo}{Paper {I}}}
\def\PaperTwo{\hyperlink{cite.London:2023idh}{Paper {II}}}

\def\ccHp#1{canonical confluent Heun polynomial#1}
\def\cHp#1{confluent Heun polynomial#1}
\def\i{(\textit{i})}
\def\ii{(\textit{ii})}
\def\iii{(\textit{iii})}
\def\ci{\textit{i}}
\def\cii{\textit{ii}}
\def\ciii{\textit{iii}}
\def\monm#1{\langle\xi^{#1}\rangle}
\def\wrt{with respect to }

\def\bpar#1{\smallskip\smallskip\paragraph*{\textbf{#1}}--~}

% Define affiliation shorthand
\newcommand{\KCL}{King's  College  London,  Strand,  London  WC2R  2LS,  United Kingdom}

% TITLE
% \title{A new special property of Schwarzschild quasinormal modes}
% \title{Properties of black hole polynomials}
\title{Properties of natural polynomials for Schwarzschild and Kerr black holes}

% AUTHOR LIST
\author{Michelle Foucoin} \affiliation{\KCL} 
\author{Lionel London} \affiliation{\KCL} 

% ABSTRACT
\begin{abstract}
The quasi-normal modes of black holes play various important roles in gravitational wave theory, signal modeling, and data analysis; however, there remain open questions about their mathematical properties. 
Aspects of classical polynomial theory have been proposed as a framework to investigate quasi-normal mode orthogonality and completeness.
We have recently presented a class of polynomials that are ``natural'' to quasi-normal modes in that they are restricted by the quasi-normal mode boundary conditions, and exactly tridiagonalize Teukolsky's radial equation.
{In turn, these polynomials may be useful for (1) better understanding the vector space properties of quasi-normal mode solutions (2) serving as a basis for spectral methods to solve the Teukolsky and confluent Heun equations.}
Here, we provide an overview of these polynomials' {basic} analytic properties: their 3-term recurrence relation, ladder operators and governing differential equation. 
We demonstrate that the natural polynomials for Schwarzschild and Kerr black holes are Pollaczek-Jacobi polynomials with complex valued parameters.
{We verify that, despite being constrained to the real line, key results for Pollaczek-Jacobi polynomials hold for black hole polynomials.}
Along the way, we observe a novel property that is particular to Schwarzschild: the polynomials' 3-term recurrence relation always peaks at the physical overtone index.    
{This work supports the broader application of these polynomials to spectral methods for solving the Teukolsky equation, and other instances of the confluent Heun equation.}
\end{abstract}

\maketitle

%%%
\section{Introduction}

\par There is now ample evidence of stellar mass \bbh{s} that merge into a perturbed remnant, which \textit{rings down}, radiating away its distortions predominantly via gravitational \qnm{s}~\cite{Kerr:1963a,TheLIGOScientific:2016src,LIGOScientific:2019fpa,LIGOScientific:2020tif,LIGOScientific:2021sio, London:2014cma,LIGOScientific:2025wao}.
The related \textit{ringdown} radiation is increasingly well understood within single \pt{}, and this understanding has been instrumental in many tests of \gr{}~\cite{Kerr:1963a,TheLIGOScientific:2016src,LIGOScientific:2019fpa,LIGOScientific:2020tif,LIGOScientific:2021sio,LIGOScientific:2025wao,LIGOScientific:2025slb}.
However, the extent to which such tests can be carried out is influenced by practical limitations in our understanding of \qnm{s}~\cite{Gupta:2024gun,Carullo:2018sfu,Carullo:2019flw,Cotesta:2022pci,Giesler:2019uxc,Isi:2019aib}.
Although numerical simulations provide high accuracy predictions of post-merger \gw{s}, our incomplete understanding of the \qnm{}s spatial completeness properties limits our ability to directly estimate their presence in simulations: 
rather than an unambiguous decomposition (i.e. projection), we must presently make heuristic and phenomenological fitting or filtering choices~\cite{London:2014cma,London:2017bcn,MaganaZertuche:2021syq,Forteza:2021wfq,Giesler:2019uxc,Giesler:2024hcr,Berti:2005ys,Berti:2016lat,Berti:2025hly,Khan:2015jqa,Husa:2015iqa,Estelles:2020osj,Ma:2023vvr,Minucci:2026dgo,Berti:2025hly}.
With future \gw{} experiments (e.g. LISA and ET) expected to require unprecedentedly complete and accurate predictions of \gr{} signals, there is good reason to refine our understanding of the \qnm{}s properties~\cite{Berti:2025hly,Klein:2015hvg,Robson:2018ifk,LISA:2022kgy,Sathyaprakash:2012jk}.
%
% With future \gw{} experiments (e.g. LISA and ET) expected to observe \rd{} signal to noise ratios of $10^3$--$10^5$, there is good reason to further refine our understanding of the \qnm{}s properties~\cite{Berti:2025hly,Klein:2015hvg,Robson:2018ifk,LISA:2022kgy,Sathyaprakash:2012jk}.
%
\par Some progress was made in Refs.~\cite{London:2023aeo} and \cite{London:2023idh}, which we will henceforth refer to as \PaperOne{} and \PaperTwo{}, respectively.
Both focus on the \qnm{s} of isolated Kerr and Schwarzschild \bh{s}, which are of known astrophysical relevance~\cite{Berti:2025hly,LIGOScientific:2025wao,LIGOScientific:2025slb}.
This is also the scope of the present work.

\par In \PaperOne{}, it was found that \qnm{} boundary conditions constrain a radial scalar product for the \qnm{s}, and that this scalar product may be used to study a restricted version of their orthogonality and completeness properties.
In \PaperTwo{}, it was demonstrated that the scalar product may be used to define polynomials that are \textit{natural} to the \qnm{} problem in the following ways:
(\textit{i}) the polynomials may be defined by applying the scalar product from \PaperOne{} to the \gs{} algorithm; therefore, their properties are tied to the \qnm{} boundary conditions, 
(\textit{ii}) the polynomials enable the representation of \qnm{} radial problem as a matrix eigenvalue problem in the simplest sense, and 
(\textit{iii}) when represented in the basis of these polynomials, the \qnm{}'s radial differential operator is exactly tridiagonal.
{\par Since the polynomials in question are natural to \bh{} solutions of \ee{}, we will henceforth refer to the class of all such functions as \textit{\bhp{s}}\footnote{
    To disambiguate between natural, \bh{}, and confluent Heun polynomials, \apx{appx:terminology} provides a summary of the terminology used in this work.
}.
While Chebyshev and Legendre polynomials are commonly used in spectral methods, the \bhp{s} offer a more tailored approach, allowing for spectral matrices that only require computation of tridiagonal elements~\cite{Chung:2023wkd,Chung:2023zdq,PanossoMacedo:2024pox,Blazquez-Salcedo:2023hwg,London:2023aeo,London:2023idh}.
Further, since the \qnm{s} radial dependence is of the confluent Heun type, the polynomials in question are also {natural} to confluent Heun problems\footnote{
    It was found in \PaperTwo{} that the radial dependence of Kerr \qnm{}s is typically well approximated by a single confluent Heun polynomial, but that the \qnm{'s} radial differential equation is generally well represented by \bhp{}s.
}. 
This implies that the \bhp{s} may be useful for other problems in physics and applied mathematics that are governed by the confluent Heun equation~\cite{ronveaux1995heun,Slavyanov:1999ep,London:2023idh,London:2023aeo,Besson2025}.}
\par In \PaperTwo{}, it was also noted that \bhp{s} for Kerr and Schwarzschild appear closely related to \pjp{s}~\cite{Chen:2010,Chen:2019,Min:2021,Min_2023}. 
A key difference would appear to be that \pjp{s} are defined on the real line, while \bhp{s} are defined on the complex plane~\cite{London:2023aeo}. 
There is therefore reason to carefully study the {basic analytic} properties of \pjp{s} in the \bh{} context.
This motivates the present work. 
\par In this paper, we present and review the {basic} properties of \bhp{s} for Schwarzschild and Kerr.
The specific properties are the \bhp{s'} governing differential equation, three-term recursion relation, and ladder operators (\textit{i.e.} derivative rule).
The main results of the present work are (1) the explicit demonstration that the analytic properties of \pjp{s} apply to \bhp{s} for Schwarzschild and Kerr, and (2) the \bhp{s} for Schwarzschild \qnm{s}, unexpectedly, maximize the average value of the radial variable; in turn, this implies a deeper connection between the \qnm{} quantization condition and the related \bh{} polynomials.

\par These findings are built upon the results of \PaperOne{}, \PaperTwo{}, and analogous results for \pjp{s}~\cite{Chen:2019,Chen:2010,Min:2021,Min_2023}.
In particular, our analysis relies heavily on Refs.~\cite{Chen:2019,Chen:2010}, which provide a semi-analytic description of the \pjp{s}' properties.
The present work assumes some basic familiarity with classical polynomial theory, and common special functions, particularly concepts such as {analytic continuation}, ladder operators and orthogonality relations.
Readers seeking background on these topics may find pedagogical introductions in Refs.~\cite{chihara2011introduction,ARFKEN2013401,VanAssche:2017,NIST:DLMF,London:2023aeo}.

\par This article is organized as follows. 
In \sec{prelims}, we provide essential background on \bh{} perturbation theory, including a review of the Teukolsky master equation and the formulation of the \qnm{} radial eigenvalue problem.
We then introduce a new choice of monomial moments for constructing \bhp{s} and demonstrate that this choice is equivalent to the construction presented in previous work, though with improved convergence properties for specific physical regimes.
In \sec{polyprops}, we establish mathematical properties satisfied by the \bhp{s}, analogous to those from classical polynomial theory, including their three-term recurrence relation, derivative rule, raising and lowering operators, and the differential equation they satisfy.
In \sec{schwarzschild}, we present the central result of this work: a special relationship between the overtone label and polynomial order in the \bhp{s} generated with respect to \qnm{} boundary conditions, particularly in the Schwarzschild limit.
Finally, in \sec{discussion}, we discuss the implications of these results for understanding \qnm{} behavior in \bh{} mergers and outline future directions for extending this framework.
%
%
%%%
\section{Preliminaries}
\label{prelims}
\par Einstein's \gr{} predicts \gw{s} that propagate with two polarizations: $h_+$ (``$h$-plus'') and $h_\times$ (``$h$-cross'') \cite{Misner1973}.
In \gw{} theory, it is useful to define a complex strain, $h$, as the double time integral of the Newman-Penrose Weyl scalar $\psi_4$ \cite{NP62,Ruiz:2007yx}.
\begin{subequations}
    \label{p1}
    \begin{align}
        h &=h_{+}-i h_{\times} 
        \\
        &=\int_{-\infty}^t \int_{-\infty}^{t^{\prime}} \psi_4\left(t^{\prime \prime}, r, \theta, \phi\right) dt^{\prime \prime}dt^{\prime},
    \end{align}
\end{subequations}
where $(t,r,\theta,\phi)$ are Boyer-Lindquist coordinates~\cite{BoyerLindquist:1967}.
In \tk{'s} formulation of linear \pt{}, a rescaling of $\psi_4$, namely,
\begin{align}
    \label{p2}
    \psi=-(r-i a \cos\theta)^{-4} \psi_4,
\end{align}
is known to solve \tk{}'s master equation, which is itself one way of representing Einstein's equations linearized about the Kerr solution \cite{Teukolsky:1973ha,Hughes:2000pf,Mino:1997bx}.
In \eqn{p2}, $a=|J|/M$ is the black hole spin \cite{Teukolsky:1973ha,Press:1971ApJ}.
In \eqn{p2}, and henceforth, it is held that $G=c=1$.
{While \eqn{p2} is specific to outgoing \grad{} (with spin weight $s=-2$), $\psi$ will generally refer to the \tk{} wave quantity which may be defined for spin weights $-2\le s \le2$ according to Ref.~\cite{Teukolsky:1973ha}.}
\par \Qnm{s} are the discrete spectra of \tk{}'s master equation.
Of the many ways to define \qnm{} solutions, the simplest begins by seeking solutions of the form 
\begin{align}
    \label{p4}
    \psi(t, r, \theta, \phi) \propto{} e^{-i \tilde{\omega} t} e^{-i m \phi} R(r) S(u),
\end{align}
{that correspond to purely ingoing waves at the \bh{} event horizon, and purely outgoing waves at spatial infinity.}
{In \eqn{p4}, } $R(r)$ and $S(u)$ are \tk{}'s radial and angular functions, respectively\footnote{Both $R(r)$ and $S(u)$ are known to be confluent Heun functions. Although $S(u)$ is often referred to as a spheroidal harmonic, the full spheroidal harmonic is $S(u)e^{i m \phi}$~\cite{Fackerell:1977,London:2020uva}.}.
\Eqn{p4} describes the functional form of a single quasinormal mode.
In \eqn{p4}, $\cw{}$ is a complex frequency,
\begin{align}
    \label{eq:qnm_freq}
    \tilde{\omega} = \omega - i/\tau,
\end{align}
where $\omega$ is the central oscillatory frequency and $1/\tau$ the positive-value exponential decay.
Each \qnm{'s} dependence on the polar angle, $\theta$, is given by a spheroidal harmonic, $S(u)$.
\par The vector space properties of the spheroidal harmonics are well understood\footnote{Various proposals for the orthogonality properties of the spheroidal harmonics are described in detail in Section 2.5 of Ref.~\cite{Berti:2025hly}.}.
The completeness of the \qnm{}'s spheroidal harmonics has been demonstrated for many physically relevant scenarios (see Ref.~\cite{London:2020uva}), and while the spheroidal harmonics are not orthogonal, there have been shown to exist dual functions (adjoint-spheroidal harmonics) with which the spheroidal harmonics are \textit{\bio{}}\footnote{
    {A sequence with elements $a_j\in A$ is \bio{} with another sequence, $b_k\in B$, if for every pair $(j,k)$ there exists a bilinear form (e.g. scalar product), $\mathcal{G}(\cdot,\cdot)$, such that $\mathcal{G}(a_j,b_k)=\delta_{jk}$.}}.
Completeness and \bio{ity} make the spheroidal harmonics compatible with the {universal} projection (i.e. decomposition) of \gw{} radiation into spheroidal multipole moments~\cite{London:2020uva,London:2021P2,Berti:2025hly}. 
\par While the vector space properties of the spheroidal harmonics $S(u)$ are well-established, properties of the radial functions, $R(r)$, remain an active area of research~\cite{London:2020uva,London:2021P2,Berti:2016lat,Minucci:2024qrn}.
We now briefly review relevant technical results from \PaperOne{} and \PaperTwo{}.
\bpar{A radial scalar product.} A radial scalar product for \qnm{s} may be constructed via Sturm-Liouville theory to the radial part of \tk{}'s master equation~\cite{London:2023aeo}. 
In particular, the product may be defined such that \tk{}'s master equation is self-adjoint (though still non-Hermitian)~\cite{London:2023aeo,Berti:2025hly,Morse:1955aqj,ARFKEN2013401}. 
We now outline how this was done in \PaperOne{}.
\smallskip
\par The radial \tk{} function, $R(r)$, satisfies an eigenvalue problem,
\begin{align}
    \label{l1}
    \mcL{_r} \, R(r) \; &= \; A \, R(r) \; ,
\end{align}
where $\mcL_r$ is a confluent Heun operator in natural general form\footnote{See Appendix C of \PaperOne{}.}.
The eigenvalue of $\mcL_r$, $A$, may be complex-valued.
It is also the separation constant for \tk{}'s master equation.
\par Following Leaver~(Ref.\cite{leaver85}), the use of a compactified radial coordinate, $\xi$, where
\begin{align}
    \label{l2}
    \xi = \frac{r - r_+}{r - r_-},
\end{align}
has the advantage of inducing polynomial coefficients in the radial operator that is derived from $\mcL_r$, once the \qnm{}'s asymmetric boundary conditions are imposed. 
This process proceeds as follows.
\par The \qnm{} boundary conditions are imposed on $R(r)$ by requiring that 
\begin{align}
    \label{l3}
    R(r(\xi)) \; = \; \mu(\xi) f(\xi)\;,
\end{align} 
and that $f(\xi)$ is finite and non-zero near the \bh{} event horizon ($\xi=0$) and space-like infinity ($\xi=1$).
Therefore, in those limits, the asymptotic behavior of $R(\xi)$ is determined by $\mu(\xi)$, which is chosen to ensure that the phase velocity of $\psi(t,r(\xi),\theta,\phi)$ is ingoing near the \bh{} event horizon and outgoing near space-like infinity.
Applying \eqns{l2}{l3} to \eqn{l1} induces a similarity transformation on the eigenvalue problem.
The result is
\begin{align}
    \label{l4}
    \mcL{_\xi} \, f(\xi) \; &= \; A \, f(\xi) \; .
\end{align}
Allowed values of $A$ are constrained by the \qnm{} boundary conditions (encapsulated by $\mu(\xi)$) and the requirement that $f(\xi)$ is analytic.
In \eqn{l4}, the radial operator, $\mcL_\xi$, is 
\begin{subequations}
    \label{l5}
    \begin{align}
        \mathcal{L}_{\xi}=&\left(\tx{C}_0+\tx{C}_1(1-\xi)\right)+\\
        &\left(\tx{C}_2+\tx{C}_3(1-\xi)+\tx{C}_4(1-\xi)^2\right)\partial_{\xi}\\
        &+\xi(\xi-1)^2 \partial_{\xi}^2 \;\; .
    \end{align}
\end{subequations}
The constants, $\tx{C}_0-\tx{C}_4$, are 
\begin{subequations}
    \label{l6}
    \begin{align}
        \label{l6a}
        \text{C}_0 \; &= \; -2 a m \cw-2 i \cw (-\delta +M(2  s+1))
        \\ \nonumber
        & \quad \;\; +\cw^2 (\delta +M) (\delta +7 M) \; ,
        \\
        \label{l6b}
        \text{C}_1 \; &= \; 8 M^2 \cw^2+ s (4 i M \cw-1)+6 i M \cw-1
        \\ \nonumber 
        & \quad \;\; - (4 M \cw+i) \frac{\left(a m-2 M^2 \cw\right)}{\delta } \; ,
    \end{align}
    \begin{align}
        \label{l6c}
        \text{C}_2 \; &= \; 4 i \delta  \cw \; ,
        \\
        \label{l6d}
        \text{C}_3 \; &= \; -2(s+1)+ 4 i \cw (M-\delta ) \; ,
        \\
        \label{l6e}
        \text{C}_4\; &= \; s+3-6 i M \cw + \frac{i \left(a m-2 M^2 \cw\right)}{\delta } \; ,
    \end{align}
    \begin{align}
        \label{l6f}
        \delta \; &= \; \sqrt{ M^2 - a^2 } \; .
    \end{align}
\end{subequations}

\par Given the transformed radial operator (\ceqn{l5}), \slt{} provides an algorithm for constructing a scalar product,
\begin{align}
    \label{l7}
    \brak{a}{b} \; = \; \int_{0}^{1} \, \mathrm{a}(\xi) \, \mathrm{b}(\xi) \; \tx{W}(\xi) \; d\xi \;,
\end{align}
under which $\mcL_\xi$ is symmetric, meaning that $\brak{a}{\mcL_\xi\, b}=\brak{\mcL_\xi \, a}{b}$~\cite{London:2023aeo}.
In \eqn{l7}, $\tx{W}(\xi)$ is called a weight function, and was found in \PaperOne{} to be 
\begin{align}
    \label{l8}
    \tx{W}(\xi) \; = \; \xi^{\tx{B}_0}(1-\xi)^{\tx{B}_1} e^{\frac{\tx{B}_2}{1-\xi}} \; ,
\end{align}
where,
\begin{subequations}
    \label{l9}
    \begin{align}
        \label{l9a}
        \text{B}_0 \; &= \; \text{C}_2+\text{C}_3+\text{C}_4-1
        \\
        \label{l9b}
        \text{B}_1 \; &= \; -\text{C}_2-\text{C}_3-2
        \\
        \label{l9c}
        \text{B}_2 \; &= \; \text{C}_2 \; .
    \end{align}
\end{subequations}
%
% {The divergence of $\tx{W}(\xi)$ as $\xi \rightarrow 1$ corresponds to the asymptotic behavior of the QNM radial function near spatial infinity (e.g. $\xi=1$ corresponds to $r \rightarrow \infty$ in the compactified coordinate). This divergence is physical given QNM solutions are outgoing at spatial infinity, with exponentially decaying imaginary part, once the relevant boundary conditions have been imposed.
% }
%
% \LTL{we can work on the working here}
%
\par {The functional form of $\tx{W}(\xi)$, including the essential singularity at $\xi=1$, reflects both the \qnm{} boundary conditions and confluent nature of \tk{'s} radial equation.
The potential for $\tx{W}(\xi)$ to diverge exponentially as $\xi \rightarrow 1$ signals that \eqn{l7} must be evaluated using analytic continuation (e.g. path integration in the complex plan, or residue calculus)~\cite{London:2023aeo}.}
\par Using this radial scalar product (\ceqn{l7}), the orthogonality and completeness properties of solutions to the radial problem were studied in \PaperOne{}.
There, it was also pointed out that the scalar product implies the existence of polynomials whose properties are tailored to \tk{}'s radial functions.
\bpar{Black hole polynomials.}
\Bhp{s} for Kerr and Schwarzschild\footnote{
    {The \qnm{s} radial equation for both Kerr and Schwarzschild \bh{s} is of the confluent Heun type. Consequently, the following discussion applies to both equally.}
} may be constructed by applying \eqns{l7}{l8} to the \gs{} algorithm\footnote{See also \PaperTwo{}, Sec.~V.}.
Henceforth, the resulting polynomials of order $k$ will be referred to as $u_k(\xi)$.
Since $\tx{W}(\xi)$ is defined in \eqn{l8} as a deformation of the (classical) Jacobi weight, $u_k(\xi)$ are \textit{semi-classical polynomials}~\cite{chihara2011introduction}.
\par In the context of the radial scalar product~(\ceqn{l7}), it will sometimes be appropriate to refer to the polynomials as bras or kets, i.e. $\bra{u_j}$ or $\ket{u_k}$, respectively.
For example, as detailed in \PaperTwo{}, \bhp{s} for Kerr and Schwarzschild are typically complete~(\ceqn{l10a}) and orthonormal~(\ceqn{l10b}):
\begin{subequations}
    \label{l10}
    \begin{align}
        \label{l10a}
        \I \; = \; \sum_{k} \ketbra{u_k}{u_k} \;,
        \\
        \label{l10b}
        \brak{u_j}{u_k} \; = \; \delta_{jk} \; .
    \end{align}
\end{subequations}
\par Concurrently, weight functions of the form of $\tx{W}(\xi)$ have been studied in applied maths, and are known to be linked to \pjp{s} (see e.g. \cite{Chen:2010,Chen:2019}).
However, the use of $\xi$~(\ceqn{l2}), as well as the normalization of the polynomials~(\ceqn{l10b}), are a departure from conventions used in {orthogonal polynomial theory}, namely Refs.~\cite{Chen:2010,Chen:2019,Min:2021,Min_2023}.
In turn, these differences impact the representation of the \pjp{} and \bhp{}'s properties.
\begin{figure*}[ht]
    
    % l  --> 2
    % m  --> 2
    % n  --> 2
    % a  --> 0.7
    % M  --> 1.0
    % s  --> -2
    % w  --> #

    \centering
    \includegraphics[width=0.49\linewidth]{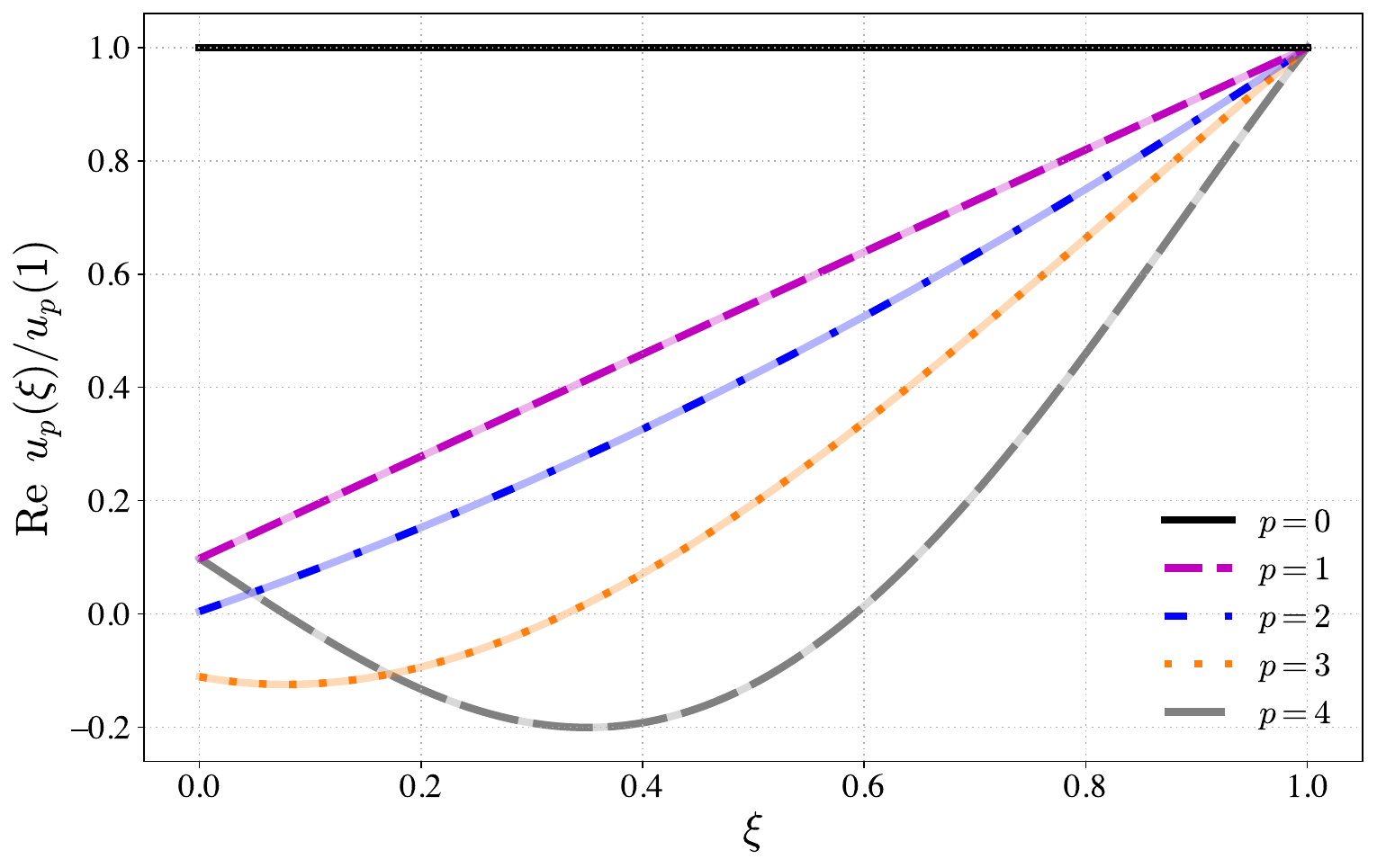}
    \hfill
    \includegraphics[width=0.49\linewidth]{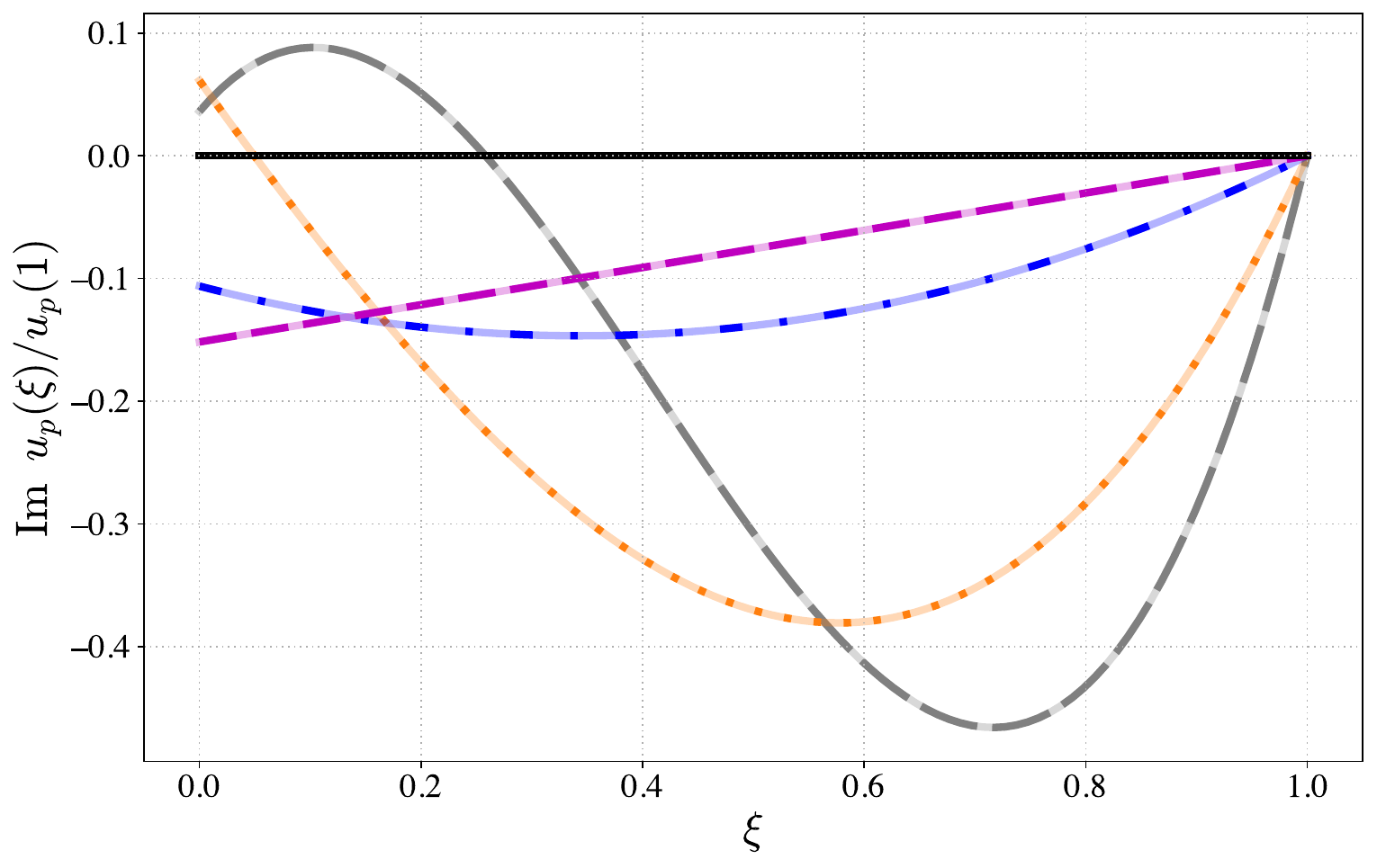}
      
    \caption{
        The first five \bhp{s}, i.e. $u_p$ for $p \in \{0, 1, 2, 3, 4\}$.
        Polynomials are constructed using spin weight $s = -2$, BH spin $a/M = 0.7$, and QNM frequency $M \bar{\omega}_{222} = 0.4999-0.4123i$.
        To make comparison more straightforward, polynomials have been normalized by their value at $\xi = 1$ (i.e. spatial infinity).
        The solid lines plot the monic orthogonal polynomials and the dashed lines the normalized polynomials, showing they are one and the same once different choices of conventions have been accounted for. 
        Left panel: Real part of each polynomial.
        Right panel: imaginary part of each polynomial.
        See discussion in \sec{prelims}.
        }
    \label{polynomials}

\end{figure*}
\bpar{Summary of conventions.}
The \bhp{s} and Pollaczek–Jacobi polynomials are closely related, but differ in the conventions chosen for their domain variables and weight functions.
To begin comparing the two sets of polynomials, it is first useful to describe how to convert between these different choices of convention.
\par The primary difference lies in the choice of domain variable: the \bhp{s}, $u_k({\xi})$, are defined with respect to $\xi \in [0,1]$, while the Pollaczek–Jacobi polynomials, $q_k(x)$, are defined with respect to $x \in [1,0]$.
The two domains are related by,
\begin{equation}
    \label{domain}
    \xi := 1-x.
\end{equation}
\Bhp{} exist within a scalar product space which uses $\tx{W}(\xi)$ from \eqn{l8},
\begin{align}
    \label{l11}
    \tx{W}(\xi)=\xi^{\tx{B}_0}(1-\xi)^{\tx{B}_1} e^{\frac{\tx{B}_2}{1-\xi}}.
\end{align}
The analogous scalar product for the \pjp{}s is
\begin{equation}
    \label{l12}
    \tx{w}(x) := x^\alpha(1-x)^\beta\mathrm{e}^{-t / x}.
\end{equation}
In \eqn{l12}, $\alpha$, $\beta$ and $t$ are typically defined as real valued parameters for which $t\ge 0$, $\alpha>0$ and $\beta>0$.
This is unlike the $\tx{B}_0$, $\tx{B}_1$ and $\tx{B}_2$ (see \ceqns{l6}{l9}) which are generally complex valued.
\par \Eqn{domain} requires the following relationship between the two weight functions,
\begin{equation}
    \label{weighfcnts}
    \tx{W}(\xi)=\tx{w}(1-x)\;.
\end{equation}
In \eqn{weighfcnts}, it has been used that a weight function may be defined up to an overall constant scale factor.
\Eqn{weighfcnts} requires that $\{\tx{B}_0,\tx{B}_1,\tx{B}_2\}$ and $\{\alpha,\beta,t\}$ are related as follows:
\begin{align}
    \label{mapping}
    \alpha = \tx{B}_1, \quad
    \beta =\tx{B}_0, \quad
    t = -\tx{B}_2.
\end{align}
\par In polynomial theory, a key use of the weight function is in computing monomial moments, which have varied uses, from deriving polynomial properties, to assessing the application of polynomials in approximation theory~\cite{chihara2011introduction,London:2023aeo}.
For \bhp{s}, the $p$-th monomial moment is defined as,
\begin{subequations}
    \begin{align}
        \label{m1a}
        \hspace{-0.365cm} \left\langle\xi^p\right\rangle & =\int_0^1 \xi^p \mathrm{~W}(\xi) d \xi \\
        \label{m1b}
        & =\int_0^1 \xi^{\mathrm{B}_0+p}(1-\xi)^{\mathrm{B}_1} e^{\frac{\mathrm{B}_2}{1-\xi}} d \xi \\
        \label{m1c}
        & =e^{\mathrm{B}_2} \Gamma\left(\mathrm{B}_0+p+1\right) U\left(\mathrm{B}_0+p+1,-\mathrm{B}_1,-\mathrm{B}_2\right).
    \end{align}
\end{subequations}
In \eqn{m1c}, $\Gamma(z)$ is the Euler Gamma function, and $U(a,b,c)$ the Tricomi confluent hypergeometric function, {both of which may be evaluated with complex valued inputs by common analytic continuation techniques (See e.g. Sec. V of \PaperOne{})}~\cite{NIST:DLMF,whittaker_watson_1996}.
\Eqn{m1c} follows from \eqn{m1b} by application of the integral form of $U(a,b,c)$.
\par For the \pjp{}s, the $p$-th monomial moment is defined as,
\begin{subequations}
    \begin{align}
        \label{m2a}
        \langle x^{p} \rangle &= \int_0^1 x^p\, \tx{w}(x)\, dx \\
        \label{m2b}
        &= \int_0^1 x^{\tx{B}_1+p}(1-x)^{\tx{B}_0} e^{\frac{\tx{B}_2}{x}}\, dx \\
        \label{m2c}
        &= e^{\tx{B}_2} \Gamma\left(\tx{B}_0+1\right) U\left(\tx{B}_0+1,\,-\tx{B}_1-p,\,-\tx{B}_2\right).
    \end{align}
\end{subequations}
In this way, the difference in domain impacts the related scalar product.
This, in turn, impacts monomial moments. 
\par Since the \bh{} and \pj{} weights differ by a change of variable,  application of the \gs{} process results in the same polynomials. 
This is as shown in \fig{polynomials}, where \bh{} and \pj{} polynomials are shown for a Kerr \bh{} with spin of $a=0.7$, and for $(\ell,m,n,s)=(2,2,2,-2)$.
There, since the values of $\tx{B}_0$, $\tx{B}_1$ and $\tx{B}_2$ are complex, so too are the related polynomials.

\bpar{Converting between monomial moments.}
Since $\xi$ and $x$ are related by a constant shift~(\ceqn{domain}), the $\left\langle \xi^p \right\rangle$ and $\left\langle x^p \right\rangle$ are related by the binomial theorem,
\begin{equation}
    \label{binomial}
    \left\langle \xi^p \right\rangle = \left\langle (1-x)^p \right\rangle = \sum_{j=0}^{p} \binom{p}{j} (-1)^j \left\langle x^j \right\rangle.
\end{equation}
In \eqn{binomial}, each $\langle \xi^p \rangle$ is expressed as a linear combination of $\langle x^j \rangle$, with binomial coefficients, $\binom{p}{j}=\frac{ p! }{ j! (p-j)! }$.
For example, $\left\langle \xi^1 \right\rangle$ may be expanded according to,
\begin{subequations}
    \label{p23}
    \begin{align}
    \left\langle \xi \right\rangle &= \left\langle 1-x \right\rangle \\
    &= \left\langle x^0 \right\rangle - \left\langle x^1 \right\rangle.
    \end{align}
\end{subequations}
The binomial theorem may also be used to write $\left\langle x^p \right\rangle$ in terms of $\left\langle \xi^p \right\rangle$, via $x=1-\xi$.
\par Since the polynomials are equivalent when evaluated with the same underlying parameters~(\ceqn{mapping}), it is reasonable to assume that they share generic properties: governing differential equation, three-term recursion relation, and ladder operators.
To verify this assumption, it is important to note a final difference in convention.

\begin{figure*}[ht]
    
    %     % l	--> 	 2
    %     % m	--> 	 2
    %     % n	--> 	 3
    %     % a	--> 	 0.0
    %     % M	--> 	 1.0
    %     % s	--> 	 -2

    \includegraphics[width=\linewidth]{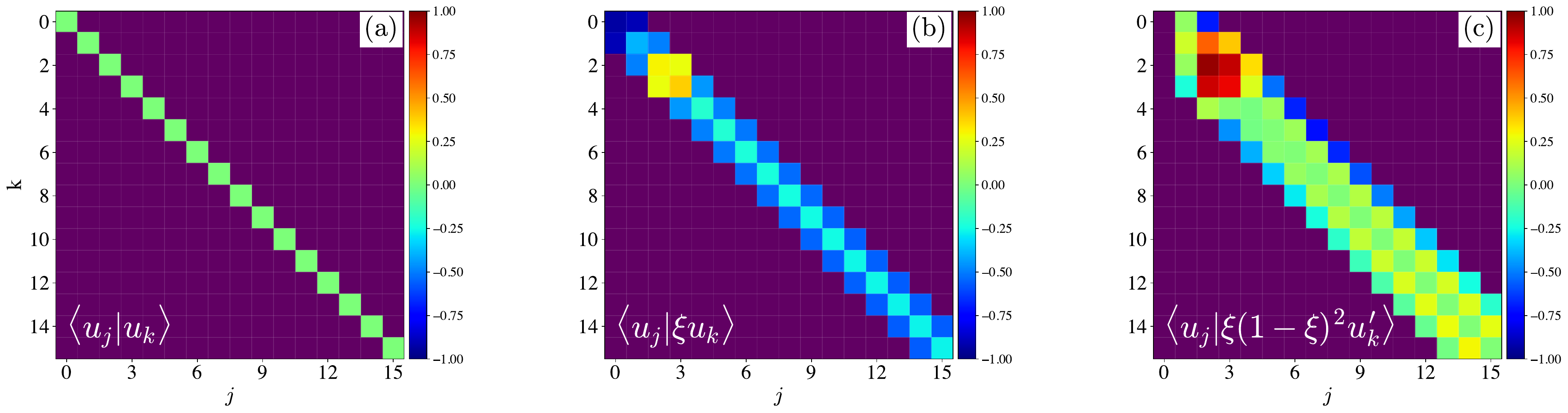} 

    \caption{ 
        Component matrices which show the scalar product calculations resulting in (a) orthogonality, (b) a three-term recurrence, and (c) a five-term recurrence are plotted on a logarithmic scale.
        All panels are computed for the $\ell, m=(2,2)$ mode of a Schwarzschild black hole with spin weight $s=-2$ and overtone $n=3$.
        Note the maximum values for panels (b) and (c) occur when overtone $n$ matches polynomial order $k$.
    } \label{gramians}

\end{figure*}

\bpar{Monic polynomials}
While \bhp{s} are typically constructed to have unit modulus~(\ceqn{l10}), meaning a norm of $1$, this is typically not the case for the \pjp{s}, whose normalization is chosen to facilitate the derivation of their analytic properties.
\par According to \eqn{l10b}, the \bhp{s} are orthonormal,
\begin{align}
    \label{eqn:normortho}
    \left\langle u_j \mid u_k \right\rangle= \int_0^1 u_{j}(\xi) u_k(\xi) \tx{W}(\xi) \, d \xi = \delta_{j,k } \; .
\end{align}
In contrast, the \pjp{}s are typically defined to be \emph{monic}, meaning that their leading coefficient is $1$,
\begin{align}
    \label{mo1}
    q_n(x)= x^n + p_1(n) x^{n-1} + \cdots.
\end{align}
In \eqn{mo1}, $q_n(x)$ is a monic \pjp{}, and $p_1(n)$ is the next to leading coefficient.
The monic polynomials are orthogonal, but not orthonormal, meaning that 
\begin{align}
    \label{mo2}
    \left\langle q_{j} \mid q_k\right\rangle=\int_0^1 q_{j}(x) q_k(x) \tx{w}(x) \, dx=h_k \,\delta_{j,k}.
\end{align}
In \eqn{mo2}, $h_k$ is the monic polynomial's normalization constant.
Consequently, the monic polynomials are related to the normalized polynomials via 
\begin{align}
    \label{mo3}
     u_{k}(x) = q_{k}(x) / \sqrt{h_k} \;,
\end{align}
where each $h_k$ is simply the $k^{th}$ entry of the diagonal Gramian matrix of the monic polynomials,
\begin{align}
    \label{mo4}
     h_k = \brak{q_k}{q_k} \; .
\end{align}
In \eqn{mo3}, and henceforth, we refer to monic \pjp{} of order $n$ as $q_{n}(x)$, and its normalized counterpart as $u_{n}(x)$.
The latter is not to be confused with $u_{n}(\xi)$, given $u_{n}(x) \neq u_{n}(\xi)$ unless $\xi=x=\frac{1}{2}$.
\par In \eqn{mo2}, the next to leading coefficient, $p_1(n)$, plays a central role in the derivation of the \pjp{s} analytic properties (see e.g. Refs.~\cite{chihara2011introduction,Chen:2010}). 
%%%
\section{Classical properties of natural polynomials for Kerr black holes}
\label{polyprops}
The semi-classical \bhp{s} share a subset of the properties of their classical counterparts (namely, orthogonality, a three-term recurrence relation, a derivative rule, and ladder operators) but do not obey a classical Rodrigues rule, and their derivatives satisfy a five-term rather than a three-term recurrence relation.
Having established in \sec{prelims} that the \bhp{s} are equivalent to the Pollaczek-Jacobi polynomials, the analytic properties derived in Refs.~\cite{Chen:2010,Chen:2019} for the latter apply directly to the former.
This section presents the following properties in detail: a three-term recurrence relation, a derivative rule and its associated ladder operators, and a second-order linear differential equation satisfied by the \bhp{s}.

\subsection{Three-term recurrence relation}
The \bhp{s} satisfy a three-term recurrence relation whose coefficients $\alpha_n$ and $\beta_n$ encode a connection between polynomial order and overtone label.
A three-term recurrence relation exists for any set of orthogonal, complete, {polynomials ordered by degree}~\cite{chihara2011introduction,VanAssche:2017}, and generally takes the form,
\begin{align}
    \label{threetermreln}
    x\, q_n(x) = q_{n+1}(x) + \alpha_n\, q_n(x) + \beta_n\, q_{n-1}(x),
\end{align}
for $n>0$. This relation follows from the fact that multiplying $q_n$ by $x$ raises its order by one, so $x\, q_n$ must be expressible as a linear combination of $q_{n+1}$, $q_n$, and $q_{n-1}$, the only three basis elements of the same or adjacent order.

The recurrence coefficient $\alpha_n$ is determined by the sub-leading coefficients of consecutive polynomials,
\begin{equation}
    \label{alphan}
    \alpha_n=p_1(n)-p_1(n+1),
\end{equation}
where $p_1(n)$ denotes the coefficient of $x^{n-1}$ in the monic polynomial, as shown in \eqn{mo1}.
% \begin{equation}
%     \label{chenpolys}
%    q_n(x) = x^n + p_1(n)\,x^{n-1} + \cdots.
% \end{equation}
%
%
\par Recall the monic polynomials $q_p$ and the orthonormal polynomials $u_p$ are related by a normalization factor, given by \eqn{mo3}, namely, $q_n = \sqrt{h_n}\, u_n$, 
%
% \begin{equation}
%     \label{qphpup}
%     q_n = \sqrt{h_n}\, u_n,
% \end{equation}
%
where $h_n = \langle q_n | q_n \rangle$ is the squared norm.
Since $\alpha_n$ arises from the ratio of leading coefficients between $q_n$ and $u_n$, it follows that,
\begin{equation}
    \label{hp}
    h_n = \alpha_n^{-2}.
\end{equation}
The coefficients $\alpha_n$ and $h_n$ thus carry information about how the polynomial normalization evolves with order, and, as shown in panel (b) of \fig{gramians}, their diagonal entries in the Gramian of the three-term recurrence peak at polynomial order $k=n$ when the polynomials are evaluated at physical \qnm{} frequencies.

\begin{figure*}[ht]

    \includegraphics[width=\linewidth]{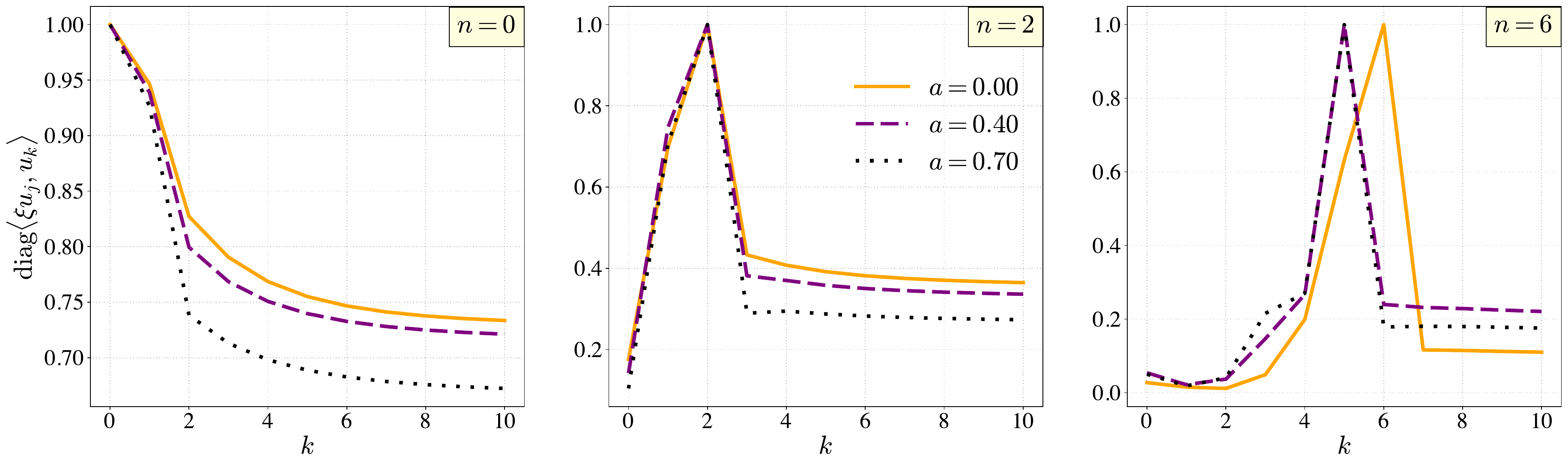} 

    \caption{ 
        The normalized three-term recurrence peaks (corresponding to the diagonal entries of the Gramian of \eqn{threetermreln}) are plotted here for the $(\ell, m)=(2,2)$ mode and spin weight $s=-2$.
        The left panel shows the peak for $n=0$, the fundamental overtone, the center panel for $n=2$, and the right panel for $n=6$.
        Notably, the peak tracks polynomial order $k$ of the corresponding recurrence relation.
        This behavior is robust to overtone number for Schwarzschild black holes, for which the peak is always observed to occur around $k=n$.
        For astrophysically relevant Kerr black holes with spin of around $a=0.40$ and $a=0.70$, the peak shifts to $k=n-1$.
    }
    \label{peaks}
\end{figure*}

%%%
\subsection{Derivative rule and ladder operators}
The \bhp{s} satisfy an analytic derivative rule that connects polynomials of consecutive order, and from which ladder operators and a five-term recurrence relation follow directly.
The derivative rule takes the form:
\begin{align}
    \label{derivativerule}
    \frac{d}{d x} q_n(x) = -V_n(x)\,q_n(x) + \beta_n U_n(x)\, q_{n-1}(x),
\end{align}
where the quantities $U_n$, $V_n$, and $\beta_n$ are defined in \apx{coeffs}, \eqns{rationalAn}{rationalBn} and \eqn{betan} respectively.
Note $\beta_n$ should not be confused with $\beta$, the weight function parameter, but rather identified as the three-term recurrence relation coefficient from \eqn{threetermreln}.
A five-term recurrence relation follows from substituting the three-term recurrence relation of \eqn{threetermreln} into the right-hand side of \eqn{derivativerule}, since each order of $q$ appearing there itself satisfies a three-term relation,
\begin{equation}
    \label{fivetermrecreln}
    \begin{split}
    x(x^2-1)\, \frac{d}{d x} q_k(x) = &c_{0,k}\, q_{k-2} + c_{1,k}\, q_{k-1} \\
    &+ c_{2,k}\, q_{k} + c_{3,k}\, q_{k+1} + c_{4,k}\, q_{k+2},
    \end{split}
\end{equation}
where the recurrence coefficients $c_{i,k}$ are calculated numerically.
As shown in panel (c) of \fig{gramians}, the Gramian of the five-term recurrence relation exhibits a similar peak structure to that of \eqn{threetermreln}, with maximum values again occurring when overtone $n$ matches polynomial order $k$.
The derivative rule of \eqn{derivativerule} may be recast as a lowering operator by rearranging,
\begin{equation}
    \boldsymbol{L}_n = \beta^{-1}_n U_n(x)^{-1} \left( \frac{d}{d x}+V_n(x) \right),
\end{equation}
such that applying $\boldsymbol{L}_n$ to a polynomial of order $n$ results in a polynomial of order $n-1$,
\begin{equation}
    \label{loweringop}
    \boldsymbol{L}_n q_n(x)=q_{n-1}(x).
\end{equation}
The corresponding raising operator, $\boldsymbol{R}_n$, follows from self-adjointness (see Ref.~\cite{VanAssche:2017}) and has the form,
\begin{equation}
    \label{raisingop}
    \boldsymbol{R}_n = U_n(x)^{-1} \left( -\frac{d}{dx} + V_n(x) + \mathcal{L}(w) \right),
\end{equation}
where $\mathcal{L}(w) = \frac{d}{d x} \ln \tx{w}(x)$ and applying $\boldsymbol{R}_n$ to a polynomial of order $n$ results in a polynomial of order $n+1$,
\begin{equation}
    \boldsymbol{R}_n q_{n}(x)=q_{n+1}(x).
\end{equation}
Since the first monic polynomial is $q_1 = 1$, repeated application of $\boldsymbol{R}_n$ generates the entire sequence and applying the raising operator $n$ times yields $q_n$.

\begin{figure*}[ht]

    \includegraphics[width=\linewidth]{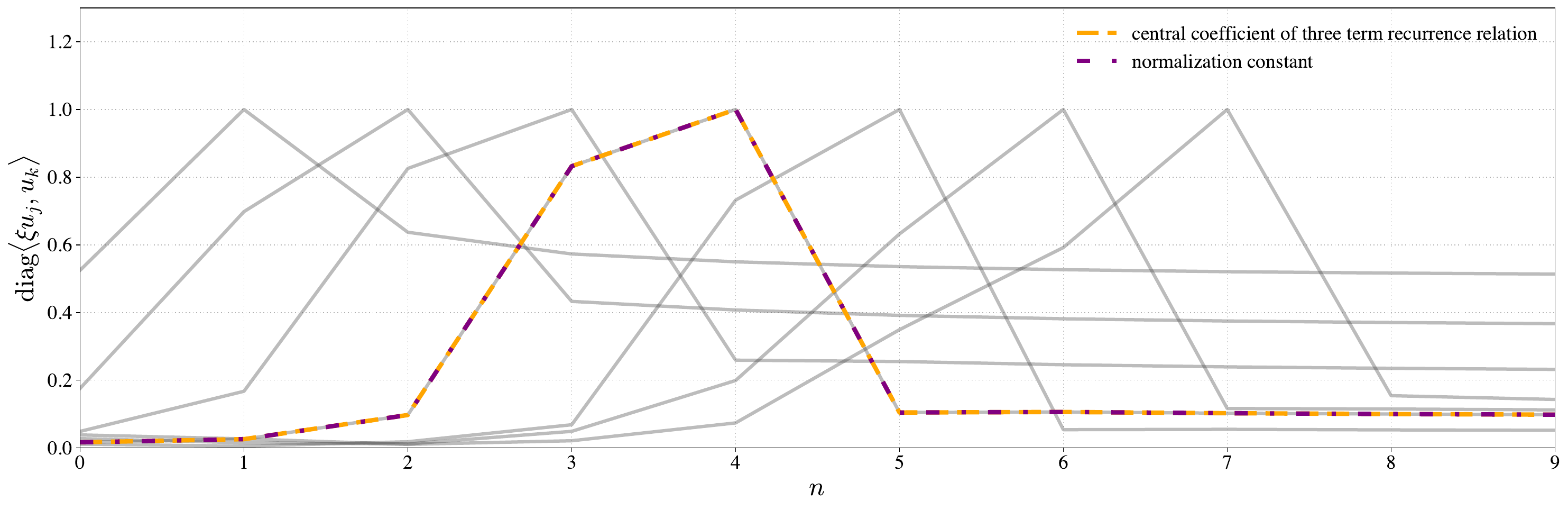} 

    \caption{ 
        Normalized peaks of the three-term recurrence relation are plotted for a sequence of overtones $n \in [1,7]$.
        All polynomials used to generate the Gramians are calculated around the Schwarzschild solution, where $a=0$, and for spin $s=-2$, and the $(2,2)$ mode.
        For each grey curve, the maximum value reached occurs at the index corresponding to overtone $n$.
        Two example quantities, $\alpha_4^{-1}$ and $\sqrt{h_4}$, are calculated and plotted for $n=4$ to show these track the peak in the same way the three-term recurrence relation does.
    }
    \label{schwpeaks}
\end{figure*}

\subsection{Differential equation}
\label{bhp-ode}
It was shown in Ref.~\cite{Min:2021} that the Pollaczek–Jacobi polynomials satisfy a linear, second-order differential equation, which, by the equivalence established in \sec{prelims}, implies \bhp{s} are also solutions.
In particular, both sets of polynomials are solutions to the homogeneous differential equation,
\begin{equation}
    \label{diffeqn}
    \frac{d^2 q_n}{dx^2} - M(x)\frac{d q_n}{dx} + N(x)\, q_n = 0
\end{equation}
where the coefficient functions are given by,
\begin{equation}
    \label{m}
    M(x) = \frac{d}{dx}\ln\left(w(x)\, U_n(x)\right)
\end{equation}
and
\begin{equation}
    \label{n}
    N(x) = U_n(x) \cdot \frac{d}{dx} \left(\frac{V_n(x)}{U_n(x)}\right) + \sum_{j=0}^{n-1} U_j(x).
\end{equation}
The quantities $U_n$ and $V_n$ are defined in \apx{coeffs}, \eqns{rationalAn}{rationalBn}, and the sum $U_j$ runs over consecutive terms.
The coefficient functions $M(x)$ and $N(x)$ are rational functions, and may be written schematically,
\begin{equation}
    \label{mschem}
    {M(x)} = \frac{m_0}{x} + \frac{m_1}{x^2} + \frac{m_2}{1-x} + \frac{m_3}{m_4 + m_5\, x},
\end{equation}
and
\begin{equation}
    \label{nschem}
    {N(x)} = \frac{n_0}{x} + \frac{n_1}{x^2} + \frac{n_2}{1-x} + \frac{n_3}{(1-x)^2},
\end{equation}
where the coefficients $m_i$ and $n_i$ are rational combinations of $\alpha$, $\beta$, $t$, and $p_1(n)$, and may be computed explicitly using the definitions of $U_n$ and $V_n$ in \apx{coeffs}, \eqns{rationalAn}{rationalBn}.

A rigorous proof that the polynomials satisfy this differential equation may be found in Ref.~\cite{Min:2021}.
As a numerical check, \eqn{diffeqn} is verified here for physically relevant cases with complex-valued coefficients, namely the outgoing $\ell,m=(2,2)$ mode of a Schwarzschild black hole, with weight function coefficients $\tx{B}_0$, $\tx{B}_1$, and $\tx{B}_2$ computed using the \texttt{positive} repository~\cite{positive:2020}.
The residual is found to be of order
\begin{equation}
    \left|\frac{d^2 q_n}{dx^2} - M(x)\frac{d q_n}{dx} + N(x)\, q_n\right| \approx 10^{-17}.
\end{equation}
Although the \bhp{s} tri-diagonalize the Teukolsky radial equation rather than solving it directly, they are themselves solutions to this different second-order linear differential equation.
Whether this equation describes a physical quantity related to linearized Kerr remains an open question.

%%%
\section{A special property for Schwarzschild}
\label{schwarzschild}
For Schwarzschild black holes, the recurrence coefficient $\alpha_n$ of the three-term recurrence relation peaks at polynomial order $k=n$ when evaluated for physical QNM frequencies, providing a direct correspondence between polynomial order and overtone label.
This section describes this property, its dependence on black hole spin, and its implications for ordering QNM solutions.

\par To understand this result, recall that the overtone label $n$ is traditionally determined by the magnitude of the imaginary part of the QNM frequency, which lacks a strict algebraic hierarchy.
The \bhp{s} offer an alternative approach to ordering overtones.
Since the \bhp{s} are generated directly from the QNM boundary conditions and tri-diagonalize the radial Teukolsky operator, their ordering emerges naturally from the physics of the problem rather than relying on an ad-hoc sorting of eigenvalues.
The observation that $\alpha_n$ peaks at $k=n$ makes this connection precise, at least in the Schwarzschild limit.

\par As shown in \fig{peaks}, the peak is clearly visible for multiple overtones in the zero spin limit, and this behavior holds across various physical combinations of $\ell,m$ indices.
Representative cases are shown in the left ($n=0$), center ($n=2$), and right ($n=6$) panels.
The inverse relationship holds (the recurrence relation achieves a \emph{minimum} rather than a maximum) for spin weight $s=+2$.
As spin increases, the peak migrates: for $a=0.40$ and $a=0.70$, it shifts to $k=n-1$ when $m=1,2$ and, in the case of $a=0.70$, the same shift occurs when $m=0$ as well.
A similar pattern holds for $\ell=3$ modes, though these are not shown.

\par The precise mechanism behind this peak, and whether an exact relationship between $k$ and $n$ can be established for general Kerr black holes, remains an open question that warrants further study.
Nevertheless, the correspondence is robust in the Schwarzschild limit and suggests the \bhp{s} may provide a more robust basis for ordering QNM solutions.

%%%
\section{Discussion and conclusions}
\label{discussion}
{Motivated by the open questions surrounding \qnm{} orthogonality and completeness, this work presents a study of the semi-classical properties of \bhp{s}.
Building on \PaperOne{} and \PaperTwo{}, which established a radial scalar product and demonstrated the existence of polynomials natural to the \qnm{} boundary value problem, the present work has shown that these \bhp{} are closely related to the semi-classical \pj{} polynomials, and that their properties follow directly from those derived in Refs.~\cite{Chen:2010,Min:2021,Min_2023,Chen:2019}.
While Refs.~\cite{Chen:2010,Min:2021,Min_2023,Chen:2019} treat \pjp{s} with real valued parameters, we have found that, despite their complex nature, the \bhp{s} satisfy the same three-term recurrence relation, derivative rule, and second-order linear differential equation as the \pj{} polynomials.
\par In \sec{prelims}, the \bhp{s} were shown to be structurally equivalent to the \pj{} polynomials, allowing their properties to be determined from those of the latter.
%
% \par In \sec{prelims}, two equivalent constructions of the \bhp{s} were presented — one based on monomial moments of $\xi$ and one on monomial moments of $x$ — and shown to generate the same polynomials up to the coordinate change $x=1-\xi$, as confirmed numerically in \fig{polynomials}.
% %
% The $\langle x^p \rangle$ moments offer improved convergence for $s=-2$ and lower overtones, while the $\langle \xi^p \rangle$ moments are preferable for $s=+2$ and higher overtones, as detailed in \apx{monomo}.
% %
% The relationship between the two normalizations is encoded in the squared norm $h_p = \alpha_p^{-2}$, where $\alpha_p$ is the recurrence coefficient of \eqn{threetermreln}, which as shown in \fig{peaks} and \fig{schwpeaks}, peaks at the overtone label $n$ when the polynomials are generated at physical \qnm{} frequencies.
%
In \sec{polyprops}, the basic semi-classical properties of the \bhp{s} were presented and verified numerically:
(\textit{i}) a three-term recurrence relation (\ceqn{threetermreln}), 
(\textit{ii}) an analytic derivative rule (\ceqn{derivativerule}) from which raising and lowering operators follow,
and (\textit{iii}) a second-order linear differential equation (\ceqn{diffeqn}) verified numerically to floating-point precision.
In all cases, we have numerically verified that the basic properties of \pjp{s} hold for physically relevant parameters of Kerr \bh{s}.
This is the main result of the present work, and it motivates further exploration of the \pjp{s} use in \bh{} physics, particularly in spectral methods in single \bh{} perturbation theory~\cite{Assaad:2025nbv,PanossoMacedo:2024pox,Zhu:2023mzv,Ripley:2022ypi,Chung:2023zdq,Chung:2023wkd,Blazquez-Salcedo:2023hwg,Besson2025}.
}
%
% The differential equation satisfied by the \bhp{s} is distinct from Teukolsky's radial equation, which the polynomials tri-diagonalize rather than solve directly; whether it describes a physical quantity related to linearized Kerr remains an open question.
%
%
% \par \LTL{Revise and update revisit last senence of abstract}
%

\begin{figure*}[ht]

    \includegraphics[width=\linewidth]{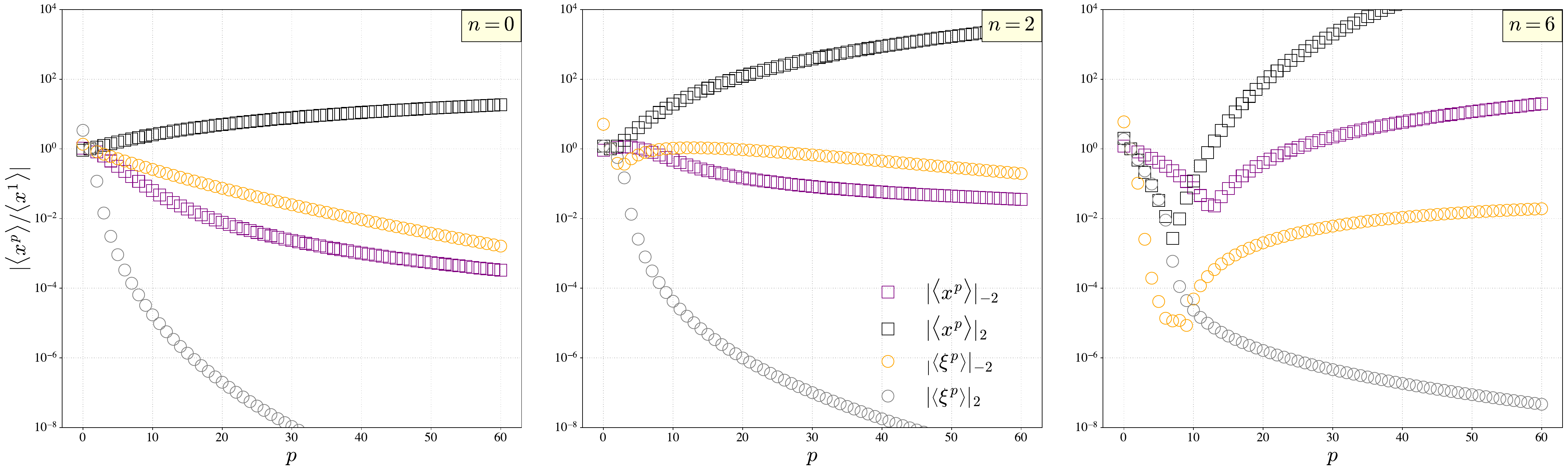} 

    \caption{ 
        Example distributions of absolute, normalized monomial moments are plotted against monomial order $p$ for the $(\ell,m)=(2,2)$ mode using the two definitions (see \eqn{m1a} and \eqn{m2a}) corresponding to two related domain variables, $\xi$ and $x$, for multiple overtones and spin weights.
        The square markers correspond to the real-valued moments and the circles to complex-valued moments, whereas the colorful markers track spin weight $s=-2$ and the greyscale $s=2$.
        The left panel shows the fundamental overtone $n=0$, the center panel $n=2$, and the right panel $n=6$.
        This is done to compare the convergence properties of the two sets of monomial moments and to study the effect of negating spin weight for either case.
        }
    \label{moments}
\end{figure*}

{
\par An {unexpected} result of this work was presented in \sec{schwarzschild}.
For Schwarzschild \bh{s}, the recurrence coefficient $\alpha_n$ peaks at polynomial order $k=n$ (i.e. the QNM overtone index) when evaluated at physical \qnm{} frequencies, as shown in \fig{peaks} and \fig{schwpeaks}.
This correspondence is robust across all combinations of $\ell$ and $m$, and persists for lower overtones at non-zero spin, though the peak migrates as $a \rightarrow 1$.
This results prompts the following question: \textit{For Schwarzschild \bh{s}, why should $\alpha_n$ know about the overtone label $n$?}
At present, we may only conjecture that this results is related to that of \PaperTwo{}, where it was found that \qnm{s} with overtone label $n$ are typically well approximated by a single confluent Heun polynomial of order $n$.
\par Developing a precise answer to that question, and the use of our \bh{} polynomials in spectral algorithms for linear perturbation theory (within and beyond GR), motivates future work.
%
% Establishing an exact relationship between overtone label $n$ and polynomial order $p$ for general Kerr black holes will be the subject of future work, as will the application of these polynomials to the study of \qnm{} completeness and their extension to alternative geometries, such as those arising from hyperboloidal slicing.
}

%%%
\section{Acknowledgements}
The authors thank Yang Chen, Alexander Pushnitski, and Jani Virtanen for supportive discussions. M. Foucoin was funded at King's College London by Royal Society enhancement grant {RF{\textbackslash}ERE{\textbackslash}210040} and by the King's College London department of physics. L. London was funded at King's College London by the Royal Society {URF{\textbackslash}R1{\textbackslash}211451}. 

% \newpage
\appendix

\section{Comparison of Monomial Moments}
\label{monomo}

The two sets of monomial moments, $\langle \xi^p \rangle$ and $\langle x^p \rangle$, differ in their asymptotic behavior and convergence properties, which has practical consequences for the numerical implementation of the \bhp{s}.
The moments exhibit different behaviors relative to spin weight $s$, for example, with fast, monotonically decreasing convergence in some instances, and slow, non-monotonic convergence in the case of $\langle x^p \rangle$ and spin weight $s=2$.
Furthermore, at higher overtones, the moments develop a non-monotonic kink near small $p$ before settling into their asymptotic trend, as seen in the $n=6$ panel.
These comparisons between the two sets of monomial moments are needed because their asymptotic behavior directly impacts the convergence properties of the sums and integrals they describe.
In both analytic and numerical contexts, whether a choice of monomial moments grows or decays (and how quickly) can slow or even prevent convergence, whereas well-controlled asymptotics allow for efficient and stable evaluation.
Understanding how different monomial sequences behave, particularly at large order, provides insight into the stability and tractability of the polynomial systems studied.

The analysis that follows is motivated by the fact that real-valued orthogonal polynomials tend to have strongly monotonic monomial moments, which in turn support rapidly convergent sums and reliable numerical behavior.
However, as was found in Ref.~\cite{London:2023aeo}, the polynomials examined in this work do not follow this typical pattern: their monomial moments display markedly different structure and asymptotics.
This deviation from the expected monotonic behavior necessitates a more careful study, both to gain a better understanding of the polynomials' analytic features and to assess numerical considerations.

The scalar product used to generate the natural polynomials, via the Gram--Schmidt algorithm, makes use of the fact that it is generally possible to describe any scalar product as a linear combination of monomial moments:
\begin{equation}
\langle\tx{a} \mid \tx{b}\rangle=\sum_{j, k} \tx{a}_j \tx{b}_k\left\langle x^{j+k}\right\rangle.
\end{equation}
The convergence of the sequences involving these moments (for example, the natural polynomials) is directly tied to the convergence of the coefficients $\tx{a}_j \tx{b}_k$.
Using the definition of \eqn{m2c}, it is possible to perform a qualitative analysis of the large order behavior in $p$.
The hypergeometric function $U$ is defined via a series expansion in its third term, and to zeroth order has an approximate form:
\begin{equation}
U(x, y, z) \approx \frac{\Gamma(1-y)}{\Gamma(x-y+1)}+\mathcal{O}(z).
\end{equation}
Combining this approximate form with \eqn{m2c} allows $\langle x^p \rangle$ to be expressed in terms of the Euler Gamma function,
\begin{equation}
    \begin{aligned}
\langle x^p \rangle & \approx \frac{\Gamma(1+\tx{B}_1+p)}{\Gamma(2+\tx{B}_0+\tx{B}_1+p)} \\
& \sim p^{-1-\tx{B}_0} \Gamma\left(1+\tx{B}_0\right).
    \end{aligned}
\end{equation}
This representation makes use of the fact that for $p \rightarrow \infty$, $\Gamma(p+\alpha) \sim p^\alpha \Gamma(p)$~\cite{abramowitz+stegun}.
The physical quantities represented by $\tx{B}_0$ may then be substituted for by using Eqns.~\eqref{l6} and \eqref{l9}, to conclude the exponent of $p$ goes as:
\begin{equation}
    \langle x^p \rangle \sim p^{-{i \left(a m-2 M^2 \omega \right)}/{\delta }+2 i M \omega +s-1}.
\end{equation}
In the zero dimensionless black hole spin limit, i.e., $a=0, M=\delta$, this simplifies to:
\begin{equation}
    \langle x^p \rangle \sim p^{-1+s + 4 i M \omega}.
\end{equation}
This behavior may be more easily compared to the asymptotics of $\langle \xi^p \rangle$ monomial moments, which obey:
\begin{equation}
    \langle\xi^p\rangle \sim p^{-1-2 s+4 i M \omega}.
\end{equation}
Indeed, in this particular limit, the behavior has very similar dependencies with a $+s$ rather than $-2s$ factor.
In general, however, the relationship is more complicated, and varies with overtone $n$.
A few example cases are plotted in \fig{moments}, for non-zero spin and varying overtone $n$, which demonstrate the two monomial moment definitions have opposite spin $s$ dependencies and converge at different rates.
In particular, the $\langle x^p \rangle$ moments grow monotonically for $s=+2$ and decay for $s=-2$, while the $\langle \xi^p \rangle$ moments exhibit the opposite spin dependence, as shown in panels (a) and (b).
Both sets develop a non-monotonic feature (see panel (c) of \fig{moments}) at higher overtones near small $p$ before settling into their asymptotic trend.

Due to the fact that the natural polynomials may be generated using either choice of monomial moments, it is important to consider the convergence rate for both.
Although the $\langle x^p \rangle$ can be generated very quickly (only $U$ and not $\Gamma$ must be calculated for each $p$), the resulting polynomial coefficients need higher numerical precision to avoid telephasing problems.
Telephasing occurs when consecutive terms in a sum nearly cancel, causing loss of significant digits and numerical instability in finite-precision calculations.
As a result, different physical scenarios call for different choices: $\langle x^p \rangle$ converges faster for $s=-2$ and lower overtone, $\langle \xi^p \rangle$ for $s=+2$ and lower overtone, and $\langle \xi^p \rangle$ for higher overtone and either spin $s$.

The behaviour of monomial moments has a direct impact on the numerical implementation of all classical polynomial properties.
When monomial moments fail to decay asymptotically, numerical evaluations of related quantities can become unstable or ill-conditioned.
To address this, extended-precision arithmetic is used to ensure reliable computation in regimes where standard precision is insufficient.
This approach is consistent with recent numerical studies, in which precision levels as high as $\texttt{dps}=400$ have been used~\cite{Besson2025}.

\section{On the analytic evaluation of \pj{} polynomials}
\label{coeffs}
Throughout this work, the results of Ref.~\cite{Chen:2010} have been used in detail.
Here, a brief practical guide to their results is provided.
\par The two main methods for generating the monic orthogonal polynomials $q_k(x)$ analytically are the three-term recurrence relation of \eqn{threetermreln} and the raising operator of \eqn{raisingop}.
Additionally, the classical properties discussed in \sec{polyprops} --- orthogonality, three-term recurrence, raising and lowering operators, and a second-order differential equation --- involve combinations of $U_n(x)$, $V_n(x)$, and $\beta_n$, which are stated here for completeness.
Together, these expressions provide a complete analytic construction of the \bhp{s}.

Both analytic constructions depend on a calculation of the logarithmic derivative of the determinant of the Hankel matrix, a matrix involving the monomial moments of the domain variable (See Sec.VI of Ref.~\cite{Chen:2010}).
{In the case of the polynomials studied, the Hankel matrix has symmetric entries consisting of consecutive monomial moments along each row and column, that is,

\begin{equation}
\mathcal{H} = \left(\begin{array}{llll}
\mu(0, \mathrm{t}) & \mu(1, \mathrm{t}) & \mu(2, \mathrm{t}) & \cdots \\
\mu(1, \mathrm{t}) & \mu(2, \mathrm{t}) & \mu(3, \mathrm{t}) & \cdots \\
\mu(2, \mathrm{t}) & \mu(3, \mathrm{t}) & \mu(4, \mathrm{t}) & \cdots \\
\vdots             & \vdots             & \vdots             & \ddots
\end{array}\right),
\end{equation}
up to a set order $n$.
The determinant of $\mathcal{H}$ is calculated in the typical way, and denoted by $D_n(t)$.
It is then possible to define an explicit form for $H_n$, the logarithmic derivative of the Hankel matrix,
\begin{equation}
H_n=t \frac{d}{d t} \ln D_n(t).
\end{equation}
}

This logarithmic derivative, $H_n$, can be expressed in terms of the weight function coefficients and the sub-leading order polynomial coefficient $p_1(n)$ of the monic polynomial,
\begin{equation}
    \label{hankeldet}
    H_n = (2n + \alpha + \beta)\,p_1(n) + n(n + \alpha - t).
\end{equation}
It is possible to solve for $p_1(n)$ by rearranging \eqn{hankeldet},
\begin{equation}
    p_1(n) = \frac{1}{\gamma_n} (H_n - n(n + \alpha - t))
\end{equation}
where $\gamma_n = 2n + \alpha + \beta$.

Recall the recurrence coefficient $\alpha_n$, which appears in the three-term recurrence relation in front of the $q_{n}$ term, was defined in \eqn{alphan} as the difference of consecutive order $p_1(n)$,
\begin{equation}
    \label{alphanappx}
    \alpha_n=p_1(n)-p_1(n+1).
\end{equation}
Therefore the Hankel determinant provides $H_n$, which in turn provides $p_1(n)$ and $\alpha_n$.
\par From this information, it is also possible to calculate the second recurrence coefficient $\beta_n$, which appears in front of the $q_{n-1}$ term,
\begin{widetext}
\begin{equation}
    \label{betan}
    \beta_n = \frac{-2p_1(n)^3 + (3t-\alpha+2\beta-2n)p_1(n)^2 - (t^2-2(n-\beta)t-(2n+\alpha)\beta)p_1(n) - (t+\beta)nt}{\gamma_n(\gamma_n^2-1) + 2p_1(n) + \gamma_n(\gamma_n-1)(\gamma_n+2)p_1(n+1) - \gamma_n(\gamma_n+1)(\gamma_n-2)p_1(n-1) - (t+\beta)(\gamma_n^2+1)}.
\end{equation}
\end{widetext}
The $q_k(x)$ follow from substituting \eqns{alphanappx}{betan} into \eqn{threetermreln} and noting $q_1=1$.
In particular, \eqn{threetermreln} may be rearranged to explicitly solve for the next polynomial in the sequence,
\begin{equation}
    q_{n+1}(x) = x\,q_n(x) - \alpha_n\,q_n(x) - \beta_n\,q_{n-1}(x).
\end{equation}

Next, consider the following equations comprised of the weight function coefficients, $\alpha$, $\beta$, and $t$, and the sub-leading order coefficients $p_1(n)$.
The goal in defining these functions is to express $U_n(x)$ and $V_n(x)$ as rational functions and offer an alternative method for calculating the $q_n(x)$ analytically.
These functions are,
\begin{equation}
    \label{Sn}
    S_n = \gamma_n\,p_1(n) - (\gamma_n + 2)\,p_1(n+1) + t + \beta
\end{equation}
and its helper function,
\begin{equation}
    \tilde{S}_n = S_n - \gamma_n - 1,
\end{equation}
as well as
\begin{equation}
    \label{Tn}
    T_n = -(p_1(n))^2 - (\gamma_n - \beta - t)\,p_1(n) + nt - \frac{(1 - \gamma_n^2)\,\beta_n}{\gamma_n},
\end{equation}
and its helper function,
\begin{equation}
    \tilde{T}_n = T_n + p_1(n).
\end{equation}
Note that $S_n$, $\tilde{S}_n$, $T_n$, and $\tilde{T}_n$ correspond to the functions $R_n$, $R_n^*$, $r_n$, and $r_n^*$, respectively, in Ref.~\cite{Chen:2010}.

\begin{table*}[htb]
    \caption{Terminology used for polynomial classes discussed in this work.}
    \label{poly-terminology}
    \renewcommand{\arraystretch}{1.3}
    \setlength{\tabcolsep}{14pt}
    \begin{tabular}{@{}p{0.20\textwidth}p{0.66\textwidth}@{}}
        \hline
        Term & Definition \\
        \hline
        Natural polynomials & Polynomials generated by applying Gram--Schmidt with the \qnm{}-motivated radial scalar product (or corresponding weight function) from the Teukolsky framework; in this manuscript they are the same object as black hole polynomials. \\
        Black hole polynomials & Solutions to the differential equation of \sec{bhp-ode} (i.e.\ \eqn{diffeqn}); equivalent to the natural polynomials of \PaperTwo{}, and constructed to exactly tridiagonalize the radial Teukolsky operator. \\
        Confluent Heun polynomials & Polynomial solutions of confluent forms of the Heun equation (special-function solutions tied to confluent Heun systems). \\
        Canonical confluent Heun polynomials & Terminology used in \PaperTwo{} for natural/black hole polynomials: they arise from confluent Heun systems while retaining canonical classical-polynomial properties (e.g.\ classical recurrence and operator structure). \\
        \hline
    \end{tabular}
\end{table*}

As a result, $U_n(x)$ and $V_n(x)$ may now be written as simple rational functions, 
\begin{equation}
    \label{rationalAn}
    U_n(x) = \frac{\tilde{S}_n}{x^2} + \frac{S_n}{x} - \frac{S_n}{x-1}
\end{equation}
and
\begin{equation}
    \label{rationalBn}
    V_n(x) = \frac{\tilde{T}_n}{x^2} - \frac{n - T_n}{x} - \frac{T_n}{x-1}.
\end{equation}
The $q_n(x)$ follow from substituting \eqns{rationalAn}{rationalBn} and \eqn{betan} into the raising operator defined in \eqn{raisingop} and noting $q_1=1$.
The rational forms of $U_n$ and $V_n$ also provide insight into why the five-term recurrence relation \eqn{fivetermrecreln} has $x(x^2-1)$ as its leading factor.
Rearranging \eqn{rationalAn} and \eqn{rationalBn}, the rational functions assume the form,
\begin{equation}
    {x^2(x-1)} U_n(x) = (\tilde{S}_n - S_n)\,x - \tilde{S}_n,
\end{equation}
and
\begin{equation}
    {x^2(x-1)} V_n(x) = (\tilde{T}_n - T_n + n)\,x - nx^2 - \tilde{T}_n.
\end{equation}
Since $U_n$ and $V_n$ appear in the derivative rule \eqn{derivativerule}, substituting these forms into the derivative rule introduces the factor $x^2(x-1)$, which is precisely the operator appearing on the left-hand side of the five-term recurrence \eqn{fivetermrecreln}.

Together, the quantities $S_n$, $\tilde{S}_n$, $T_n$, $\tilde{T}_n$, and $\beta_n$ provide a complete and efficient characterization of the \bhp{s}' derivative rule and recurrence structure, expressing all relevant quantities as rational functions of $x$ with coefficients determined entirely by the weight function parameters and the Hankel determinant.
If the parameter mapping \eqn{mapping} is applied, these terms depend instead on the coefficients of \eqn{l9}, which are determined by combinations of the physical parameters of the system: $\delta$, $M$, $s$, and $\tilde{\omega}$.

\section{Heun, confluent Heun, and black hole polynomials}
\label{appx:terminology}

{
\par Here and in \tbl{poly-terminology} we clarify the terminology used in this work and in \PaperTwo{} regarding Heun polynomials, confluent Heun polynomials, and black hole polynomials. 
\par Heun polynomials are solutions of the Heun equation, and confluent Heun polynomials are solutions of its confluent form~\cite{ronveaux1995heun}. 
The ``black hole polynomials'' are solutions to the differential equation of \sec{bhp-ode}, i.e. \eqn{diffeqn}.
Black hole polynomials arise from application of the Gram--Schmidt algorithm to a weight function from the \tk{} equation, and as such are precisely the ``natural polynomials'' studied in \PaperTwo{}.
\par \PaperTwo{} also describes the natural polynomials as \textit{canonical confluent Heun polynomials}, since they may be derived from confluent Heun systems while retaining the canonical properties of classical polynomials.
In practice, the use of black hole polynomials (equivalently, natural polynomials) to represent the radial \tk{} operator results in an exactly tridiagonal form, whereas confluent Heun polynomials do not.
}

\bibliography{references.bib}

\end{document}